\documentclass[sigconf,nonacm]{acmart}
\usepackage[utf8]{inputenc}
\usepackage{textgreek}
\usepackage{xcolor}
\usepackage{amsmath}
\usepackage[ruled, vlined, linesnumbered]{algorithm2e}
\usepackage{bm}
\usepackage{enumitem}
\usepackage{lipsum}
\usepackage[most]{tcolorbox}
\usepackage{tikz}
\usepackage{pgfplots}
\usepgfplotslibrary{groupplots}
\pgfplotsset{compat=1.18}
\usetikzlibrary{patterns}
\usetikzlibrary{arrows.meta,calc,decorations.pathreplacing,shapes.geometric}
\usepackage[aboveskip=-4pt]{subcaption}
\usepgfplotslibrary{fillbetween}
\usepackage{amsthm}
\usepackage{mdframed}
\usepackage{seqsplit}
\usepackage{placeins}
\usepackage{multirow}
\usepackage{colortbl}

\usepackage{booktabs}
\usepackage{array}
\usepackage{adjustbox}
\usepackage{tabularx}
\usepackage{array}
\newcolumntype{R}{>{\raggedleft\arraybackslash}X}

\theoremstyle{definition}

\newcommand{\cc}[1]{{\color{blue}{#1}}}

\newcommand{\stitle}[1]{\vspace*{0.2em}\noindent{\bf #1.\/}}

\definecolor{hlbg}{RGB}{200,240,240}

\newcommand{\ourbench}{\textsc{CHO-Bench}\xspace}
\newcommand{\ourproblem}{\textsc{CHO}\xspace}

\newcommand{\adlib}{\textit{AdLib-US}\xspace}
\newcommand{\adindia}{\textit{AdLib-IN}\xspace}
\newcommand{\naiad}{\textit{NaiAD}\xspace}
\newcommand{\lmmarket}{\textit{LM-Market}\xspace}
\newcommand{\egeo}{\textit{E-GEO}\xspace}

\newcommand{\db}{\mathbf{d}}

\newcommand{\pb}{\mathbf{p}}

\newcommand{\xb}{\mathbf{x}}

\def\addlegendimage{\csname pgfplots@addlegendimage\endcsname}

\makeatletter
\newcommand\footnoteref[1]{\protected@xdef\@thefnmark{\ref{#1}}\@footnotemark}
\makeatother

\definecolor{RedPPAlt}{HTML}{DC143C}
\definecolor{Red}{HTML}{E81123}
\definecolor{Orange}{HTML}{FFB000}
\definecolor{Green}{HTML}{009E49}
\definecolor{LightBlue}{HTML}{00BCF2}
\definecolor{DeepBlue}{HTML}{001BA3}
\definecolor{Pink}{HTML}{7C3AED}
\definecolor{LightGreen}{HTML}{8BC63E}
\definecolor{Gray}{HTML}{6B7280}

\newcounter{myremark}
\renewcommand{\themyremark}{\arabic{myremark}}

\newtcolorbox{remark}[1][]{
    enhanced,
    breakable,
    before skip=10pt,
    after skip=10pt,
    colback=gray!5,
    colframe=navyblue!70!black,
    boxrule=0.3pt,
    leftrule=3pt,
    left=8pt,
    right=5pt,
    top=5pt,
    bottom=5pt,
    sharp corners,
    fonttitle=\bfseries,
    coltitle=black,
    title={Remark \refstepcounter{myremark}\themyremark #1},
    attach title to upper,
    after title={.\hspace{0.5em}},
}
\definecolor{navyblue}{RGB}{0, 0, 128}

\newcommand{\trref}[1]{Appendix~\ref{appendix}}

\AtBeginDocument{%
  }

\begin{document}

\title{\textsc{CHO-Bench}: How Do Context Hints Affect Ad Rankings in LLM
Advertising?}

\author{Shiqi Zhang}
\orcid{0000-0002-7155-9579}
\affiliation{%
  \institution{National University of Singapore}
  \country{Singapore}
}
\affiliation{%
  \institution{PyroWis AI}
  \country{Singapore}
}
\email{shiqi@pyrowis.ai}

\author{Yang Fei}
\affiliation{%
  \institution{National University of Singapore}
  \country{Singapore}
}
\email{yfei11@u.nus.edu}

\author{Xiaokui Xiao}
\orcid{0000-0003-0914-4580}
\affiliation{%
  \institution{National University of Singapore}
  \country{Singapore}
}
\email{xkxiao@nus.edu.sg}

\begin{abstract}
LLM advertising selects advertisements based on user conversations, where user needs may extend beyond explicit product searches. ChatGPT Ads allows advertisers to provide natural-language \textit{context hints} describing what they offer, who may find it useful, and when it may be useful, as additional information for ranking advertisements across diverse user conversations. However, how to effectively construct these hints remains unclear. We formulate Context Hint Optimization (\ourproblem) as the problem of constructing context hints to improve an advertisement's ranking under a given budget. Drawing on established marketing frameworks, we develop seven context hint generation strategies covering What, Who, and When, along with a Unified strategy that combines all seven dimensions. We introduce \ourbench and conduct over 1.19 million paired evaluations across five datasets using embedding-based and LLM-based rankers. Our results show that context hints improve advertisement rankings under the evaluated offline rankers. While individual strategies achieve similar average gains and Unified offers no consistent advantage, strategy differences become more pronounced under specific generation--ranking configurations, with embedding-based and LLM-based rankers favoring different information dimensions. Increasing the hint budget further improves rankings with diminishing marginal gains. These findings establish an offline benchmark for studying ad-side ranking optimization in LLM advertising and highlight opportunities for model-specific hint selection and budget-aware generation and aggregation.
\end{abstract}

\maketitle

\section{Introduction}
\label{sec:intro}

\begin{figure}[!t]
\centering
\def\adpath{fig/ads}
\newdimen\adtopdim
\newsavebox\adbox
\definecolor{tCook}{HTML}{2E9D6A}
\definecolor{tStep}{HTML}{D99A1E}
\definecolor{tLook}{HTML}{8A5CC2}
\definecolor{tNutr}{HTML}{3B6EA8}
\definecolor{tFit}{HTML}{C8285E}
\colorlet{cHL}{tFit}
\definecolor{cBlueLink}{HTML}{2F7FD8}
\definecolor{cEye}{HTML}{5CD3F0}
\definecolor{cTag}{HTML}{3F8F3A}
\newcommand{\adphoto}[5]{%
  \begin{scope}
    \clip[rounded corners=3pt] (#2,#3) rectangle (#4,#5);
    \IfFileExists{\adpath/#1}{%
      \sbox\adbox{\includegraphics{\adpath/#1}}%
      \pgfmathsetmacro{\phw}{(#4)-(#2)}\pgfmathsetmacro{\phh}{(#5)-(#3)}%
      \pgfmathsetmacro{\imgr}{\the\wd\adbox/\the\ht\adbox}%
      \pgfmathparse{\imgr > \phw/\phh ? 1 : 0}%
      \ifnum\pgfmathresult=1
        \node[inner sep=0pt] at ($(#2,#3)!0.5!(#4,#5)$) {\includegraphics[height=\phh cm]{\adpath/#1}};
      \else
        \node[inner sep=0pt] at ($(#2,#3)!0.5!(#4,#5)$) {\includegraphics[width=\phw cm]{\adpath/#1}};
      \fi
    }{%
      \fill[black!8] (#2,#3) rectangle (#4,#5);
      \node[font=\sffamily\scriptsize, text=black!40, align=center] at ($(#2,#3)!0.5!(#4,#5)$) {photo\\#1};
    }%
  \end{scope}}
\tikzset{
  pics/robot/.style={code={
    \draw[#1!75!black, line width=0.5pt] (0,0.3) -- (0,0.4);
    \fill[#1] (0,0.42) circle (0.05);
    \fill[#1] (-0.32,0.08) rectangle (-0.27,0.22);
    \fill[#1] (0.27,0.08) rectangle (0.32,0.22);
    \draw[#1!75!black, fill=white, line width=0.6pt, rounded corners=2.5pt] (-0.27,0) rectangle (0.27,0.3);
    \fill[#1!25!black, rounded corners=1.5pt] (-0.2,0.05) rectangle (0.2,0.25);
    \fill[cEye] (-0.085,0.15) ellipse (0.045 and 0.055);
    \fill[cEye] (0.085,0.15) ellipse (0.045 and 0.055);
    \draw[#1!75!black, fill=#1!45, line width=0.5pt, rounded corners=2pt] (-0.19,-0.25) rectangle (0.19,-0.03);
    \draw[#1!75!black, fill=white, line width=0.4pt] (0,-0.14) circle (0.055);
  }},
  pics/usericon/.style={code={
    \fill[#1!15, rounded corners=3pt] (-0.22,-0.22) rectangle (0.22,0.22);
    \fill[#1] (0,0.06) circle (0.075);
    \fill[#1] (-0.13,-0.15) arc (180:0:0.13 and 0.12) -- cycle;
  }},
  pics/pot/.style={code={
    \draw[#1, line width=0.8pt, rounded corners=2pt] (-0.17,0.06) -- (-0.17,-0.15) -- (0.17,-0.15) -- (0.17,0.06);
    \draw[#1, line width=0.8pt] (-0.22,0.06) -- (0.22,0.06);
    \draw[#1, line width=0.8pt] (-0.12,0.1) arc (180:0:0.12 and 0.06);
    \fill[#1] (0,0.18) circle (0.03);
    \draw[#1, line width=0.8pt] (-0.17,-0.02) -- (-0.24,-0.02) (0.17,-0.02) -- (0.24,-0.02);
  }},
  pics/steps/.style={code={
    \foreach \i in {0,1,2} {
      \fill[#1] (-0.17,{0.13-0.13*\i}) circle (0.035);
      \draw[#1, line width=0.8pt, line cap=round] (-0.08,{0.13-0.13*\i}) -- (0.2,{0.13-0.13*\i});
    }
  }},
  pics/recipe/.style={code={
    \draw[#1, line width=0.8pt, rounded corners=1.5pt] (-0.14,-0.2) rectangle (0.14,0.2);
    \foreach \y in {0.08,0,-0.08} \draw[#1, line width=0.6pt] (-0.07,\y) -- (0.07,\y);
    \draw[#1, line width=0.8pt, fill=white] (0.12,-0.1) circle (0.07);
    \draw[#1, line width=0.9pt, line cap=round] (0.17,-0.15) -- (0.22,-0.2);
  }},
  pics/chat/.style={code={
    \draw[#1, line width=0.8pt] (0,0.02) ellipse (0.21 and 0.16);
    \fill[white] (-0.13,-0.11) -- (-0.21,-0.21) -- (-0.04,-0.14) -- cycle;
    \draw[#1, line width=0.8pt] (-0.12,-0.105) -- (-0.21,-0.21) -- (-0.05,-0.14);
    \foreach \x in {-0.08,0,0.08} \fill[#1] (\x,0.02) circle (0.025);
  }},
  pics/dumbbell/.style={code={
    \draw[#1, line width=1.2pt] (-0.12,0) -- (0.12,0);
    \fill[#1, rounded corners=1pt] (-0.19,-0.13) rectangle (-0.12,0.13);
    \fill[#1, rounded corners=1pt] (0.12,-0.13) rectangle (0.19,0.13);
    \fill[#1, rounded corners=1pt] (-0.25,-0.08) rectangle (-0.2,0.08);
    \fill[#1, rounded corners=1pt] (0.2,-0.08) rectangle (0.25,0.08);
  }},
}
\resizebox{\columnwidth}{!}{%
\begin{tikzpicture}[
  font=\sffamily\small,
  qcard/.style={fill=#1!7, draw=#1!30, line width=0.5pt, rounded corners=4pt},
  acard/.style={fill=white, draw=black!15, line width=0.5pt, rounded corners=4pt},
  hl/.style={draw=cHL, line width=1.3pt, rounded corners=4pt},
  vdot/.style={circle, fill=#1, inner sep=0pt, minimum size=0.16cm},
  halo/.style={circle, fill=cHL!25, inner sep=0pt, minimum size=0.36cm},
  pill/.style={fill=#1!12, text=#1!65!black, rounded corners=4pt, font=\sffamily\footnotesize,
               inner xsep=4pt, inner ysep=1.5pt, text height=1.6ex, text depth=0.45ex},
]
\def\ct{7.15} \def\cb{0.4} \def\hy{7.52}

\fill[black!3, rounded corners=6pt] (0,0) rectangle (6.0,7.9);
\node[font=\sffamily\small\bfseries] at (1.55,\hy) {User queries};
\node[font=\sffamily\small\bfseries] at (5.0,\hy)  {Displayed ads};

\foreach \n/\i/\c/\ic/\topic/\q in {
    q1/0/tCook/pot/Cooking Tips/{``How to cook tofu?''},
    q2/1/tStep/steps/How-to Guide/{``Give me a step-by-step guide to make pad Thai.''},
    q3/2/tFit/dumbbell/Fitness Goal/{``How can I build muscle?''},
    q4/3/tNutr/chat/Nutrition Q\&A/{``Why do people say pasta is `empty' calories?''},
    q5/4/tLook/recipe/Recipe Lookup/{``Recipe for lasagna.''}} {
  \pgfmathsetmacro{\y}{\ct-0.6-1.3875*\i}
  \draw[qcard=\c] (0.15,\y-0.6) rectangle (2.95,\y+0.6);
  \pic[scale=0.95] at (0.45,\y) {\ic=\c};
  \node[anchor=north west, inner sep=0pt, text=\c!85!black, font=\sffamily\footnotesize\bfseries]
    at (0.78,\y+0.47) {\topic};
  \node[anchor=north west, inner sep=0pt, text width=2.05cm, align=left,
        font=\sffamily\footnotesize, execute at begin node={\hyphenpenalty=10000}, text=black!85]
    at (0.78,\y+0.21) {\q};
  \coordinate (\n) at (3.03,\y);
}
\draw[hl] (0.15,{\ct-0.6-1.3875*2-0.6}) rectangle (2.95,{\ct-0.6-1.3875*2+0.6});

\foreach \n/\i/\brand/\file in {a1/0/Factor/factor.jpg, a2/1/HelloFresh/hellofresh.jpg,
                                a3/2/Peloton/peloton.jpg, a4/3/Scheels/scheels.jpg} {
  \pgfmathsetmacro{\y}{\ct-0.75-1.75*\i}
  \draw[acard] (4.15,\y-0.75) rectangle (5.85,\y+0.75);
  \node[anchor=west, inner sep=0pt, font=\sffamily\footnotesize\bfseries, text=black!80]
    at (4.27,\y+0.56) {\brand};
  \adphoto{\file}{4.25}{\y-0.65}{5.75}{\y+0.35}
  \coordinate (\n) at (4.07,\y);
}
\draw[hl] (4.15,{\ct-0.75-1.75-0.75}) rectangle (5.85,{\ct-0.75-1.75+0.75});

\foreach \q in {q1,q2,q4,q5} \draw[line width=0.7pt, cBlueLink!40] (\q) -- (a2);
\foreach \a in {a1,a3,a4}     \draw[line width=0.7pt, cHL!40] (q3) -- (\a);
\draw[line width=3.4pt, cHL!22, line cap=round] (q3) -- (a2);
\draw[line width=1.4pt, cHL] (q3) -- (a2);
\foreach \q/\c in {q1/tCook,q2/tStep,q4/tNutr,q5/tLook} \node[vdot=\c] at (\q) {};
\foreach \a in {a1,a3,a4} \node[vdot=black!35] at (\a) {};
\foreach \p in {q3,a2} { \node[halo] at (\p) {}; \node[vdot=cHL] at (\p) {}; }

\fill[cHL!3, draw=cHL!25, line width=0.5pt, rounded corners=6pt] (6.2,0) rectangle (12.25,7.9);
\begin{scope}
  \clip[rounded corners=6pt] (6.2,0) rectangle (12.25,7.9);
  \fill[cHL!12] (6.2,7.15) rectangle (12.25,7.9);
\end{scope}
\node[anchor=west, font=\sffamily\footnotesize\bfseries] at (6.35,\hy)
  {Example: ``How can I build muscle?'' $\rightarrow$ \textcolor{cBlueLink}{HelloFresh}};

\pic at (6.57,6.7) {usericon=cBlueLink};
\node[anchor=west, fill=cBlueLink!10, rounded corners=4pt, inner xsep=5pt, inner ysep=3pt,
      font=\sffamily\footnotesize] at (6.92,6.7) {How can I build muscle?};

\pic[scale=0.85] at (6.57,6.0) {robot=tCook};
\node[anchor=north west, inner sep=0pt, text width=5.05cm, align=left, font=\sffamily\footnotesize,
      execute at begin node={\hyphenpenalty=10000}]
  (r0) at (6.95,6.2)
  {Building muscle comes down to three pillars: \textbf{training}, \textbf{nutrition}, and \textbf{recovery}.};
\node[anchor=north west, inner sep=1.5pt, text width=4.95cm, align=left, font=\sffamily\footnotesize]
  (r1) at ($(r0.south west)+(-0.02,-0.08)$)
  {\textbf{1.\ Strength training:} progressive overload with compound lifts.};
\node[anchor=north west, inner sep=1.5pt, text width=4.95cm, align=left, font=\sffamily\footnotesize,
      fill=cHL!10, rounded corners=2pt]
  (r2) at ($(r1.south west)+(0,-0.05)$)
  {\textbf{2.\ Nutrition:} 0.7--1\,g protein per lb of body weight and a slight calorie surplus.};
\node[anchor=north west, inner sep=1.5pt, text width=4.95cm, align=left, font=\sffamily\footnotesize]
  (r3) at ($(r2.south west)+(0,-0.05)$)
  {\textbf{3.\ Recovery:} 7--9\,h of sleep and rest days.\,\ldots};

\path (r3.south) ++(0,-0.2) coordinate (brTop);
\node[anchor=north west, draw=cHL!60, dashed, fill=cHL!5, rounded corners=4pt, line width=0.6pt,
      text width=5.57cm, align=left, inner sep=4pt, font=\sffamily\footnotesize, text=black!80]
  (br) at (6.35,0|-brTop)
  {This conversation discusses \textcolor{cHL}{\textbf{nutrition and meal planning}} for muscle gain,
   which is \textcolor{cHL}{\textbf{relevant to meal kit services like HelloFresh.}}};

\path (br.south) ++(0,-0.18) coordinate (adTop);
\pgfextracty{\adtopdim}{\pgfpointanchor{adTop}{center}}
\pgfmathsetmacro{\adtopy}{\adtopdim/28.45274}
\draw[acard] (6.35,\cb) rectangle (12.1,\adtopy);
\IfFileExists{\adpath/hellofresh_card.jpg}
  {\adphoto{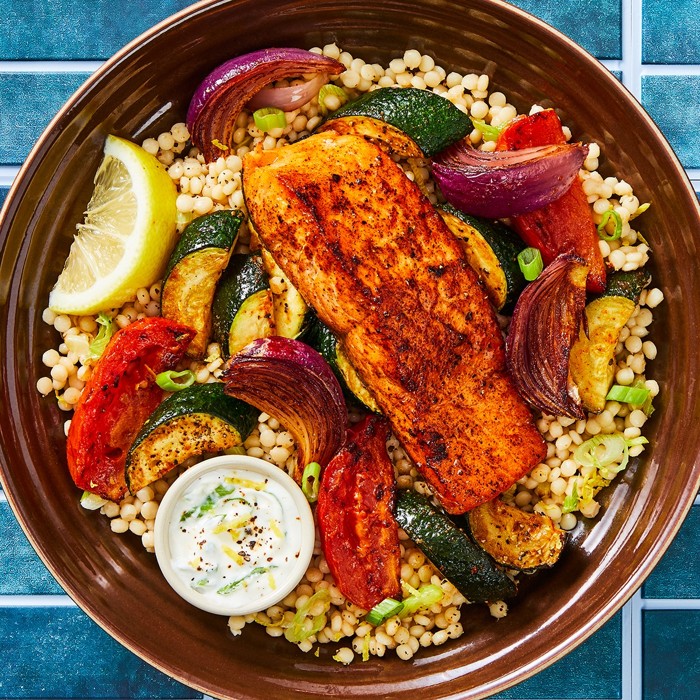}{10.2}{\cb+0.12}{11.98}{\adtopy-0.12}}
  {\adphoto{hellofresh.jpg}{10.2}{\cb+0.12}{11.98}{\adtopy-0.12}}
\begin{scope}
  \clip (6.62,\adtopy-0.33) circle (0.18);
  \IfFileExists{\adpath/hellofresh.jpg}
    {\node[inner sep=0pt] at (6.62,\adtopy-0.33) {\includegraphics[height=0.36cm]{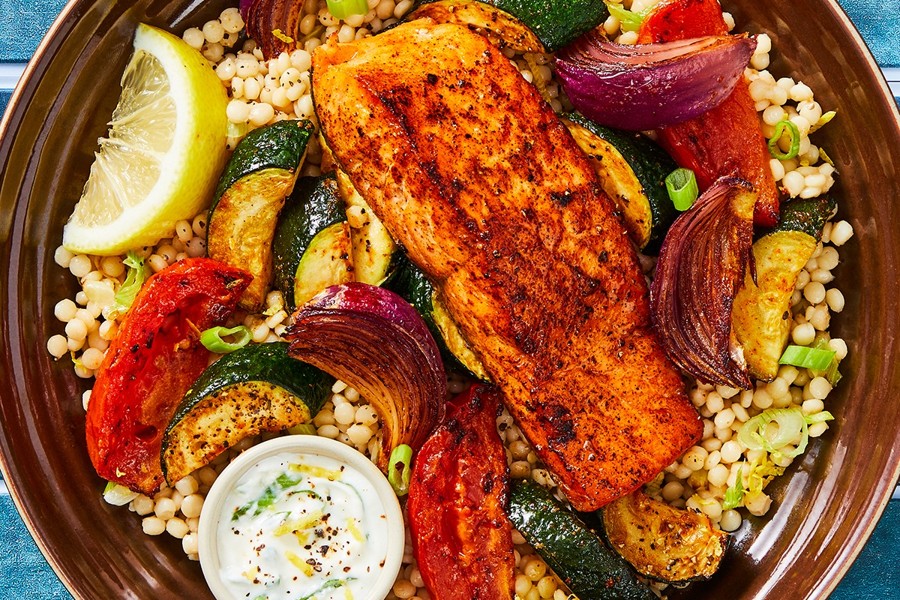}};}
    {\fill[cTag!25] (6.62,\adtopy-0.33) circle (0.18);}
\end{scope}
\node[anchor=west, inner sep=0pt, font=\sffamily\small\bfseries] at (6.88,\adtopy-0.25) {HelloFresh};
\node[anchor=west, inner sep=0pt, font=\sffamily\scriptsize, text=black!50] at (6.88,\adtopy-0.55) {Sponsored};
\node[anchor=north west, inner sep=0pt, font=\sffamily\footnotesize\bfseries] (adt) at (6.47,\adtopy-0.82)
  {Try Our Protein Smart Meals};
\node[anchor=north west, inner sep=0pt, text width=3.6cm, align=left, font=\sffamily\footnotesize, text=black!70]
  at ($(adt.south west)+(0,-0.08)$) {30g+ protein per serving.};
\node[pill=cTag, anchor=south west] (p1) at (6.47,\cb+0.12) {Meal kit};
\node[pill=cTag, anchor=south west] at ($(p1.south east)+(0.08,0)$) {High-protein};

\end{tikzpicture}}
\vspace{-4mm}
\caption{Real example in ChatGPT~\citep{lurie2026beginning}. Left: Queries and advertisements displayed in ChatGPT's responses. Right: ChatGPT's response to ``How can I build muscle?'' and the displayed advertisement for high-protein meals.}
\label{fig:motivation}
\vspace{-4mm}
\end{figure}
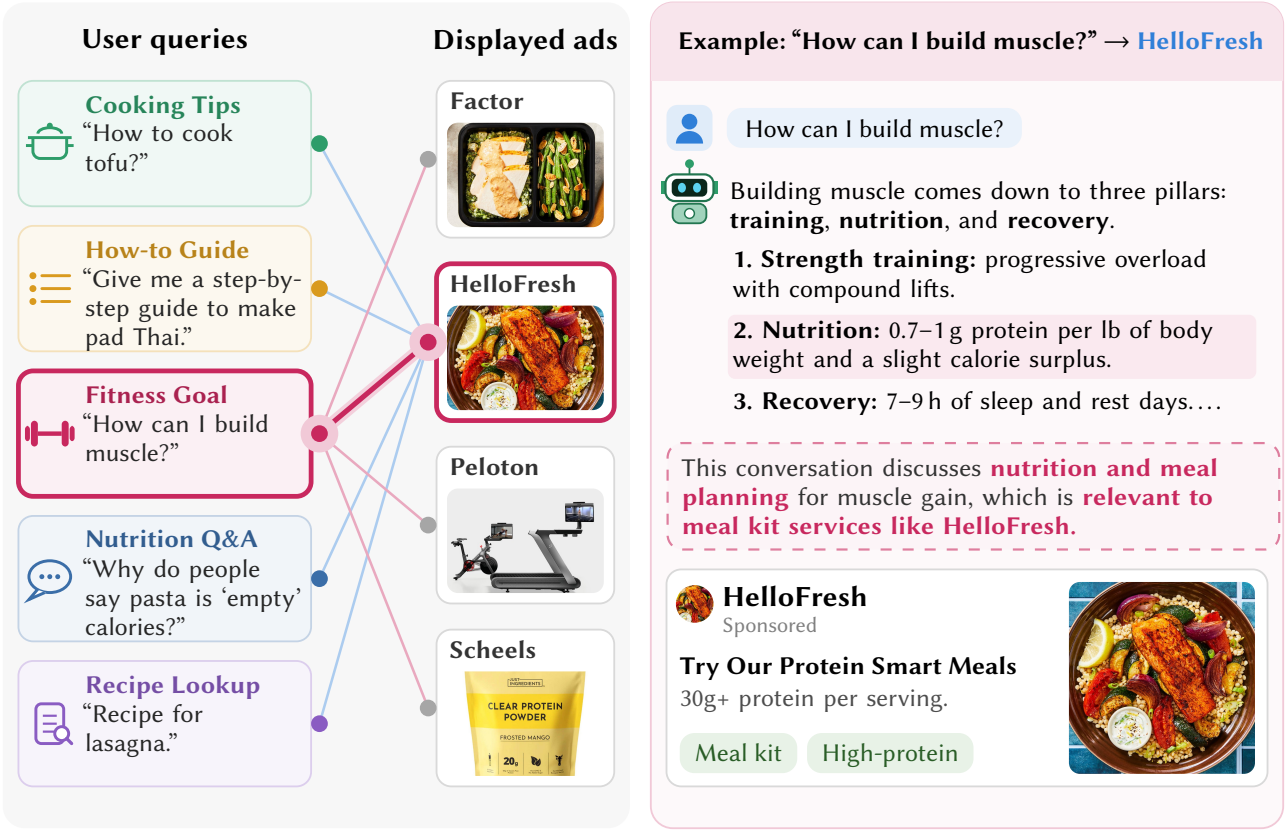

Large language model (LLM) assistants and generative engines are emerging as new advertising channels. Platforms including ChatGPT, Microsoft Copilot, and Google AI Search now integrate sponsored content into conversational or AI-generated experiences~\citep{openai_chatgpt_ads,microsoft_copilot_ads,google_ai_overviews_ads}. LLM advertising selects advertisements based on user conversations and presents them alongside AI-generated responses~\citep{duetting2024mechanism,hajiaghayi2024ad,hu2026gem}.
A key challenge for advertisers in LLM advertising is reaching users across diverse conversational contexts. Beyond explicit product searches, users may seek information, solve problems, or pursue goals without directly requesting a product or service. A recent empirical study of ChatGPT Ads reported non-obvious connections between user queries and displayed advertisements~\citep{lurie2026beginning}. As illustrated in Figure~\ref{fig:motivation}, for the query ``How can I build muscle?'', ChatGPT displayed advertisements for workout services, fitness equipment, and meal delivery. Conversely, HelloFresh, a meal-kit delivery service, appeared across queries involving cooking instructions, recipes, nutrition, and fitness. Although the muscle-building query does not directly request meal delivery, ChatGPT's response discusses protein intake, providing a connection to HelloFresh's Protein Smart meals. This example illustrates how the same advertisement can be associated with diverse user needs and situations.

In ChatGPT Ads, advertisers can provide free-form natural-language \textit{context hints} describing what they offer, who may find it useful, and when it may be useful~\citep{openai_chatgpt_ads}. These hints supplement ad cards with information about potential user needs and situations, allowing advertisements to express connections to conversations beyond explicit product searches. Unlike optimizing ranking models, ad creatives, or external web content, context hints offer an additional input for ad ranking. However, whether context hints improve advertisement rankings and how to construct them effectively remain unclear.
We formulate \textit{Context Hint Optimization} (\ourproblem) as the problem of constructing context hints to improve an advertisement's ranking under a given budget. This formulation raises three questions. Can context hints improve advertisement rankings? Which information dimensions and combinations are most effective? How does their effectiveness vary across ranking models and hint budgets?

To address these questions, we first develop context hint generation strategies. Following the \textit{What}, \textit{Who}, and \textit{When} categories in ChatGPT Ads guidance, we draw on established marketing frameworks, including the Value Proposition Canvas~\citep{osterwalder2014value} and Customer Journey~\citep{lemon2016customer}, to identify seven information dimensions. These dimensions cover product features, customer needs, and usage situations. We develop a generation strategy for each dimension and introduce a Unified strategy that combines information across all seven dimensions. Together, these strategies provide a basis for comparing information selection and aggregation.

To evaluate \ourproblem across different ranking systems, we consider two fundamental approaches to ad ranking, namely embedding-based similarity scoring~\citep{grbovic2016scalable,hu2026gem,zhang2026naiad} and LLM-based listwise ranking~\citep{sun2023chatgpt,zhuang2024setwise}. Both serve as building blocks for more complex ranking pipelines. Based on these approaches, we introduce \ourbench to evaluate context hints under different ranking mechanisms.
Since \ourproblem optimizes context hints rather than ad cards or ranking models, \ourbench compares an advertisement's ranking with and without context hints for the same query and candidate pool, while keeping other advertisements, ad cards, and ranking functions fixed. This paired design measures the incremental ranking effect of context hints and enables reproducible offline evaluation without access to proprietary advertising systems.

Using \ourbench, we evaluate eight context hint generation strategies across five datasets, two generation models, and four ranking models, yielding 320 experimental configurations. We further evaluate Unified under five hint budgets, adding 200 configurations. Across three repetitions, our experiments comprise 520 configurations and over 1.19 million paired evaluations.
Our experiments reveal three main findings. First, context hints improve advertisement rankings under the evaluated offline rankers, achieving gains of 0.56--1.16 positions in $\Delta$ Rank and 5.42--14.40 percentage points in $\Delta$ Win across general and product advertising datasets.
Second, the seven information strategies achieve similar average gains across datasets, with differences of 0.07--0.24 ranking positions, and Unified does not consistently outperform individual strategies. However, these differences can reach 0.57 positions under specific generation--ranking configurations, with embedding-based and LLM-based rankers favoring different information dimensions.
Third, increasing the hint budget further improves advertisement rankings with diminishing marginal gains, particularly for embedding-based rankers. These findings highlight opportunities for model-specific information selection and budget-aware hint generation and aggregation.

Our contributions are summarized as follows:
\begin{itemize}[topsep=2pt,itemsep=1pt,parsep=0pt,partopsep=0pt,leftmargin=11pt]
    \item We formulate \ourproblem as an ad-side ranking optimization problem and develop eight marketing-grounded strategies.

    \item We introduce \ourbench, an offline benchmark using embedding and LLM rankers with controlled paired evaluations.

    \item We conduct over 1.19 million paired evaluations, revealing the effects of hint strategies, ranking models, and hint budgets.
\end{itemize}

\section{Preliminaries}
\label{sec:prelim}

\subsection{Problem Formulation}
\label{sec:problem}

LLM advertising extends online advertising to LLM-based assistants. Following prior work~\citep{duetting2024mechanism,hajiaghayi2024ad,hu2026gem}, we describe the advertising pipeline in four stages.
\begin{itemize}[topsep=2pt,itemsep=1pt,parsep=0pt,partopsep=0pt,leftmargin=11pt]
    \item \textbf{Campaign configuration.}
    Advertisers provide campaign objectives, budgets, bids, targeting settings, ad creatives, and other information for ad delivery.

    \item \textbf{Ad ranking.}
    Given a user interaction, the platform identifies and ranks candidate advertisements based on the conversational context and available ad information.

    \item \textbf{Auction and allocation.}
    Eligible advertisements are selected through an allocation mechanism that may consider relevance, bids, and other platform objectives.

    \item \textbf{Ad presentation.}
    Selected advertisements are displayed within or alongside AI-generated responses.
\end{itemize}

\stitle{ChatGPT Ads}
ChatGPT Ads Manager provides a concrete instance of this pipeline~\citep{openai_chatgpt_ads}. During campaign configuration, advertisers specify campaign settings $\gamma$, including objectives, budgets, and bids. Each advertisement $a$ contains a user-facing ad card $\mathcal{C}_a$, comprising a title, description, and creative, and a set of natural-language context hints $\mathcal{H}_a$ describing what the advertisement offers, who may find it useful, and when it may be useful. Ads Manager can also collect information from the advertiser's website, which we denote as the ad corpus $\mathcal{D}_a$, to initialize the ad card and context hints. Both components can subsequently be revised.
During ad ranking, ChatGPT uses the conversation context, ad cards, and context hints to rank candidate advertisements. Eligible advertisements then enter a relevance-weighted second-price auction, after which selected ad cards are displayed below ChatGPT responses. Unlike ad cards, which contain the information presented to users, context hints provide additional information for ranking advertisements based on user conversations.

\stitle{Context Hint Optimization}
We focus on optimizing context hints while keeping other components of the advertising system unchanged. Specifically, for an advertisement $a$, we fix its ad card $\mathcal{C}_a$, campaign settings $\gamma$, and the ad-ranking function $f$. Given a hint budget $k$, \textit{Context Hint Optimization} (\ourproblem) aims to construct a set of context hints $\mathcal{H}_a$ that maximizes the advertisement's ranking utility:
\begin{equation}\label{eq:obj}
\mathcal{H}_a^*
=
\arg\max_{\mathcal{H}_a \in \mathcal{F}_a,\,
|\mathcal{H}_a|\leq k}
U_f(\mathcal{C}_a,\mathcal{H}_a),
\end{equation}
where $\mathcal{F}_a$ denotes the feasible sets of context hints for advertisement $a$, and $U_f(\mathcal{C}_a,\mathcal{H}_a)$ denotes its expected ranking utility across user conversations under the fixed ranking function $f$. The utility can be instantiated using objectives such as negative average rank or top-1 placement rate. Context hints may be constructed from advertiser information or external sources. In this work, we restrict hint generation to the ad corpus $\mathcal{D}_a$ to compare information strategies under a common information source.

\subsection{Related Work}
\label{sec:related}

\stitle{LLM advertising}
Existing research investigates different stages of the LLM advertising pipeline, including bidding, mechanism design, and ad allocation~\citep{duetting2024mechanism,hajiaghayi2024ad,dubey2024auctions}, as well as ad ranking, selection, and integration into LLM-generated responses~\citep{tang2024genai,hu2026gem,schmidt2024detecting}. Recent empirical studies further examine real-world ad delivery and user outcomes~\citep{lurie2026beginning}. Our work introduces context hint optimization as an ad-side ranking optimization problem that keeps ranking, auction, and presentation mechanisms fixed.

\stitle{Ad ranking}
Ad ranking aims to rank advertisements for user queries and contexts. Existing methods employ semantic representations of queries and advertisements~\citep{grbovic2016scalable} and incorporate user behaviors, preferences, and personalized relevance signals~\citep{hong2024print,zhang2024scaling}. More recently, \citet{mohankumar2024improving} leverage web search results and LLM-generated query explanations to improve sponsored search retrieval. While these approaches improve ranking functions $f$, we keep $f$ fixed and optimize context hints $\mathcal{H}_a$ as an additional input to improve advertisement rankings.

\stitle{Advertising creative optimization}
Advertising creative optimization focuses on the content presented to users. Prior studies generate and optimize textual advertisements~\citep{mishra2020learning,mita2024striking,zhang2024adtec} and visual or multimodal creatives~\citep{chen2021automated,yang2024creative} to improve engagement, click-through rates, and conversions. In contrast, \ourproblem keeps the user-facing ad card $\mathcal{C}_a$ unchanged and optimizes context hints $\mathcal{H}_a$ used for ad ranking.

\stitle{Generative engine optimization}
Generative engine optimization (GEO) aims to improve the visibility of content in LLM-generated responses. Existing methods rewrite or restructure external content~\citep{aggarwal2024geo,c-seo,autogeo}, while others study adversarial content optimization to influence generated responses and rankings~\citep{ragdoll,black-hat-geo,tang2025stealthrank}. Both GEO and \ourproblem optimize information supplied to LLM-based systems, but differ in their optimization targets. GEO modifies content consumed by generative engines to influence its visibility in generated responses, whereas \ourproblem optimizes context hints $\mathcal{H}_a$ to improve advertisement rankings while keeping ad cards $\mathcal{C}_a$ and ranking functions $f$ fixed.

\section{Data Collection and Preprocessing}
\label{sec:data}

We construct \ourbench from five datasets covering two advertising settings: \textit{general ads}, spanning diverse conversational needs and advertisement types, and \textit{product ads}, focusing on product recommendations. For general ads, we use \adlib~\citep{lurie2026beginning} and \adindia~\citep{viren2026chatgptads,trucommerce2026chatgptads} from public ChatGPT ad libraries, together with \lmmarket and \naiad from existing LLM advertising benchmarks~\citep{hu2026gem,zhang2026naiad}. For product ads, we use \egeo~\citep{bagga2025egeo}. We construct candidate ad pools using observed query--ad associations or semantic retrieval and prepare an ad card $\mathcal{C}_a$ and ad corpus $\mathcal{D}_a$ for each advertisement. All ad cards follow the ChatGPT Ads limits of 50 characters for titles and 100 for descriptions. Table~\ref{tab:data} summarizes the datasets, with additional preprocessing details in~\trref{}.

\stitle{\adlib and \adindia}
We construct two datasets from public libraries of advertisements displayed in ChatGPT. \adlib is derived from \citet{lurie2026beginning}, which recorded 3,602 advertisement observations across 127,801 conversations in the U.S. market. We group advertisements by query, remove duplicates, and retain queries with advertisements from at least two brands, yielding 109 queries, 883 ads, and 1,093 observed query--ad pairs. \adindia combines two libraries primarily covering the Indian market~\citep{viren2026chatgptads,trucommerce2026chatgptads}. After filtering incomplete queries and advertisements, we group prompts expressing the same need and supplement candidate pools containing fewer than 10 brands with semantically similar ads, following \citet{hu2026gem}. This yields 72 queries, 67 ads, and 740 query--ad pairs. For both datasets, we retain the original ad cards and construct ad corpora through LLM-assisted web search of advertiser landing pages.

\stitle{\lmmarket and \naiad}
We derive \lmmarket from marketing-related prompts in \citet{hu2026gem} and \naiad from real user queries in \citet{zhang2026naiad}. Two LLM judges independently assess query suitability for advertising, followed by manual verification, yielding 387 and 243 queries, respectively. Neither dataset provides observed ChatGPT ad placements, so we retrieve candidate advertisements from a shared library of 4,778 ads across 610 topics~\citep{hu2026gem,tang2024genai}. Following \citet{hu2026gem,zhang2026naiad}, we select the 10 most semantically similar advertisements from distinct brands for each query, yielding 3,870 and 2,430 query--ad pairs. We construct ad cards from advertiser names and descriptions and obtain ad corpora through the same LLM-assisted web-search procedure.

\begin{table}[!t]
\centering
\small
\caption{Summary of the five preprocessed datasets. }
\label{tab:data}
\vspace{-2mm}
\setlength{\tabcolsep}{5pt}
\renewcommand{\arraystretch}{1.12}

\begin{tabular}{llrrr}
\toprule
\textbf{Setting}
& \textbf{Dataset}
& \textbf{\# Queries}
& \textbf{\# Ads}
& \textbf{\# Pairs} \\

\midrule

General
& \adindia~\citep{trucommerce2026chatgptads,viren2026chatgptads}
& 72 & 67 & 740 \\

General
& \adlib~\citep{lurie2026beginning}
& 109 & 883 & 1,093 \\

General
& \naiad~\citep{zhang2026naiad}
& 243 & 1,155 & 2,430 \\

General
& \lmmarket~\citep{hu2026gem}
& 387 & 1,389 & 3,870 \\

Product
& \egeo~\citep{bagga2025egeo}
& 1,934 & 16,446 & 19,316 \\

\bottomrule
\end{tabular}
\vspace{-4mm}
\end{table}

\stitle{\egeo}
We use \egeo~\citep{bagga2025egeo}, which contains shopping queries adapted from Reddit product recommendation requests and their retrieved Amazon products. From its held-out test set of 2,000 queries, we retain advertising-suitable queries with candidate products from at least two brands, yielding 1,934 queries, 16,446 products, and 19,316 query--product pairs. Unlike the general-ad datasets, \egeo provides structured product metadata. We use this metadata directly as the ad corpus $\mathcal{D}_a$ and construct length-constrained ad cards $\mathcal{C}_a$ by extracting product identities and key selling points using an LLM.

\section{\ourproblem Strategies}
\label{sec:framework}

OpenAI recommends that context hints describe \textit{what} an advertiser offers, \textit{who} may find it relevant, and \textit{when} it may be useful~\citep{openai_chatgpt_ads}. These categories provide general guidance but do not specify how to select information for context hints. To identify specific information dimensions, we draw on two established marketing frameworks. The Value Proposition Canvas (VPC)~\citep{osterwalder2014value} describes an offering's value through its Value Map and customer needs through its Customer Profile. The Customer Journey~\citep{lemon2016customer} describes how customers interact with offerings across different decision stages. Based on these frameworks, we identify seven information dimensions and develop a corresponding context hint generation strategy for each. Figure~\ref{fig:pipeline} illustrates the strategies using the HelloFresh Protein Smart meals advertisement displayed for ``How can I build muscle?'' in Figure~\ref{fig:motivation}.

\input{fig/pipeline}

\subsection{What Dimension}

The \textit{what} dimension describes the offering and the value it provides. Following the Value Map of the VPC~\citep{osterwalder2014value}, we consider three strategies corresponding to products and services, pain relievers, and gain creators.

\stitle{Product Features}
This strategy captures factual information about an offering, including its features, capabilities, services, specifications, pricing, and availability, rather than customer outcomes. For HelloFresh, the collector identifies Protein Smart meals containing at least 30g of protein per serving.

\stitle{Pain Relievers}
This strategy identifies problems, risks, frustrations, or obstacles that an offering can help reduce or resolve. Each hint links a customer problem to a stated product feature or service. For HelloFresh, meal kits can reduce the effort of planning and preparing protein-rich meals.

\stitle{Gain Creators}
This strategy identifies desirable customer outcomes supported by an offering's features or services. For HelloFresh, Protein Smart meals can help customers maintain a convenient, protein-rich eating routine.

\subsection{Who Dimension}

The \textit{who} dimension identifies customers who may be interested in an offering. Based on the Customer Profile in VPC~\citep{osterwalder2014value}, we distinguish Customer Segments, which describe customer groups, from Customer Jobs, which describe underlying tasks and goals.

\stitle{Customer Segments}
This strategy identifies customer groups based on their roles, backgrounds, characteristics, or circumstances, rather than their immediate tasks. For HelloFresh, the collector identifies regular exercisers, people following higher-protein diets, and those seeking to build or maintain muscle. The latter connects to the muscle-building query in Figure~\ref{fig:motivation}, even though the query does not mention meals or nutrition.

\stitle{Customer Jobs}
Drawing on Jobs-to-be-Done (JTBD)~\citep{christensen2016competing}, this strategy identifies practical tasks and functional, social, or emotional goals that customers seek to accomplish. It emphasizes underlying customer purposes rather than purchasing the offering itself. For HelloFresh, these include meeting protein targets and supporting fitness or strength goals through meal choices.

\subsection{When Dimension}

The \textit{when} dimension describes the circumstances in which an offering may be useful. Drawing on the Customer Journey~\citep{lemon2016customer}, we distinguish Usage Scenarios, which describe situations creating a need for the offering, from Decision Stages, which describe customers' progress toward a purchase decision.

\stitle{Usage Scenarios}
This strategy identifies events, activities, or circumstances that make an offering useful, rather than general goals or purchase decisions. For HelloFresh, examples include planning meals for a higher-protein eating routine and choosing protein-rich meals after strength training.

\stitle{Decision Stages}
This strategy captures information needs at different stages of the customer journey. During \textit{awareness}, customers recognize a problem or goal without seeking a specific offering. During \textit{consideration}, they explore or compare solutions. During \textit{purchase}, they choose, order, or book a solution. For HelloFresh, these stages correspond to exploring higher-protein meal options, comparing meals by protein content, and deciding whether to order Protein Smart meals.

\section{\ourbench}
\label{sec:benchmark}

\subsection{Benchmark Pipeline}
\label{sec:benchmark-pipeline}

Figure~\ref{fig:pipeline} illustrates the two-stage pipeline of \ourbench. In the first stage, \textit{Hint Collectors} extract information from the ad corpus $\mathcal{D}_a$, and a \textit{Hint Aggregator} generates context hints $\mathcal{H}_a$ for each advertisement $a$. In the second stage, an \textit{Ad Ranker} ranks advertisements for each query, and a \textit{Rank Evaluator} measures changes in their rankings. Ad cards remain fixed throughout the evaluation.

\stitle{Hint Collectors}
Given an ad corpus $\mathcal{D}_a$, each collector generates candidate statements for one of the seven information dimensions in Section~\ref{sec:framework}, without a predefined limit on the number of statements. The statements describe specific product attributes, customer needs, or situations, avoiding generic or repetitive content. Product claims must be grounded in $\mathcal{D}_a$, while reasonable inferences about customer needs, goals, and situations are allowed. Each dimension-specific generation strategy uses its corresponding collector, while Unified combines statements from all seven collectors. Collectors do not access evaluation queries or ranking outcomes.

\stitle{Hint Aggregator}
A shared aggregator selects and combines candidate statements into context hints $\mathcal{H}_a$, following OpenAI's context hint guidelines~\citep{openai_chatgpt_ads}. When a hint budget is specified, the aggregator follows the corresponding constraint. The resulting hints describe potential connections between an offering and user needs or situations. In Figure~\ref{fig:pipeline}, the aggregator consolidates 23 candidate statements about HelloFresh Protein Smart meals into five hints covering protein intake, meal planning, and fitness-related needs. All generation strategies use the same aggregator and differ only in the information supplied by their collectors.

\stitle{Ad Ranker}
We instantiate the ranking function $f$ in Eq.~\eqref{eq:obj} using embedding-based similarity scoring and LLM-based listwise ranking. Each ranker ranks advertisements within a fixed candidate pool using their ad cards $\mathcal{C}_a$ and available context hints $\mathcal{H}_a$. The resulting rankings provide the basis for evaluating the ranking effects of context hints.

\stitle{Rank Evaluator}
The rank evaluator compares advertisement rankings with and without context hints under a fixed ranking procedure and candidate pool.

\subsection{Evaluation Protocol}
\label{sec:evaluation-protocol}

\stitle{Paired Evaluation}
For each query, we select one or more advertisements from its candidate pool as \textit{targets}, where each target is an advertisement whose context hints are evaluated. Each query--target pair defines a benchmark instance $i$, consisting of a query $q$, target advertisement $a$, and candidate ad pool $\mathcal{A}_q$. We compare a control condition without context hints against a treatment condition that adds hints generated by the evaluated strategy to the target advertisement. Both conditions use the same query, candidate pool, and ad cards, while all non-target advertisements remain unchanged. The ranking function $f$ produces rankings $\pi_i$ and $\pi_i'$ under control and treatment, respectively.
We generate context hints three times with different random seeds and average the results across repetitions. For LLM-based ranking, we evaluate each instance four times with different candidate orders, using the same four orders for control and treatment.

\stitle{Target Selection}
For \adlib and \adindia, we use advertisements observed in ChatGPT responses as targets. In \adlib, each observed query--ad pair defines a benchmark instance. In \adindia, advertisements observed for any query within the same group serve as targets for all queries in that group, excluding advertisements added through semantic retrieval. For \lmmarket and \naiad, we randomly select one target advertisement from each query's candidate pool. For \egeo, we use the target product specified by the original dataset. These procedures yield 1,093, 175, 387, 243, and 1,934 benchmark instances, respectively. Target selection determines which advertisements are evaluated and does not imply independently verified query--ad relevance.

\begin{table*}[t]
\centering
\caption{Effects of context hint generation strategies across five datasets. No Hints reports baseline ranking performance, while other strategies report ranking gains relative to the baseline. Dimension results average the strategies within each dimension. Relative Ranking reports the mean rank of each strategy or dimension across experimental settings (mean $\pm$ SD, lower is better). All aggregated strategy gains are statistically significant (paired $t$-tests with standard errors clustered by target advertisement and Holm-adjusted $p<0.001$).}
\label{tab:rq1-main}

\vspace{-1.5mm}
\small
\setlength{\tabcolsep}{1.45pt}
\renewcommand{\arraystretch}{1.12}

\begin{tabular}{llrrrrrrrrrrrr}
\toprule
\multicolumn{2}{c}{}
& \multicolumn{2}{c}{\textbf{\adlib}}
& \multicolumn{2}{c}{\textbf{\adindia}}
& \multicolumn{2}{c}{\textbf{\lmmarket}}
& \multicolumn{2}{c}{\textbf{\naiad}}
& \multicolumn{2}{c}{\textbf{\egeo}}
& \multicolumn{2}{c}{\textbf{Relative Ranking $\pm$ SD}} \\
\cmidrule(lr){3-4}
\cmidrule(lr){5-6}
\cmidrule(lr){7-8}
\cmidrule(lr){9-10}
\cmidrule(lr){11-12}
\cmidrule(l){13-14}

\textbf{Dimension} & \textbf{Strategy}
& $\Delta$ Rank & $\Delta$ Win
& $\Delta$ Rank & $\Delta$ Win
& $\Delta$ Rank & $\Delta$ Win
& $\Delta$ Rank & $\Delta$ Win
& $\Delta$ Rank & $\Delta$ Win
& $\Delta$ Rank & $\Delta$ Win\\
\midrule

-- & No Hints
& 7.28 & 17.82
& 4.79 & 19.23
& 5.28 & 11.68
& 5.43 & 9.20
& 5.16 & 12.44
& -- & -- \\

\midrule
Who & Customer Segments
& +0.93 & +6.01
& +0.80 & +7.50
& +0.94 & +9.23
& +1.06 & +10.70
& +1.11 & +13.90
& 3.67 $\pm$ 2.04 & 3.55 $\pm$ 2.07 \\

& Customer Jobs
& +0.96 & +5.98
& +0.61 & +6.05
& +0.96 & +9.42
& +1.07 & +10.98
& +1.06 & +13.45
& 4.21 $\pm$ 2.20 & 4.11 $\pm$ 2.13 \\

& \textit{Dimension mean}
& +0.94 & +5.99
& +0.71 & +6.78
& +0.95 & +9.33
& +1.07 & +10.84
& +1.08 & +13.68
& 1.65 $\pm$ 0.70 & 1.68 $\pm$ 0.73 \\

\midrule
What & Product Features
& +0.86 & +5.61
& +0.63 & +6.58
& +0.90 & +8.77
& +1.01 & +9.88
& +1.06 & +13.29
& 5.80 $\pm$ 1.47 & 5.47 $\pm$ 1.94 \\

& Pain Relievers
& +0.84 & +5.49
& +0.56 & +6.07
& +0.90 & +8.94
& +0.97 & +9.26
& +1.08 & +13.80
& 5.94 $\pm$ 2.09 & 5.49 $\pm$ 2.16 \\

& Gain Creators
& +0.87 & +5.42
& +0.62 & +6.72
& +0.92 & +8.73
& +1.02 & +10.01
& +1.06 & +13.46
& 5.31 $\pm$ 1.86 & 5.29 $\pm$ 2.01 \\

& \textit{Dimension mean}
& +0.86 & +5.50
& +0.61 & +6.46
& +0.91 & +8.81
& +1.00 & +9.72
& +1.07 & +13.52
& 2.52 $\pm$ 0.78 & 2.30 $\pm$ 0.85 \\

\midrule
When & Usage Scenarios
& +0.94 & +5.46
& +0.68 & +6.70
& +0.97 & +9.45
& +1.09 & +11.15
& +1.05 & +13.16
& 4.25 $\pm$ 2.33 & 4.99 $\pm$ 2.51 \\

& Decision Stages
& +0.87 & +5.80
& +0.74 & +6.96
& +0.97 & +9.39
& +1.01 & +9.74
& +1.16 & +14.40
& 4.26 $\pm$ 2.47 & 4.16 $\pm$ 2.28 \\

& \textit{Dimension mean}
& +0.91 & +5.63
& +0.71 & +6.83
& +0.97 & +9.42
& +1.05 & +10.45
& +1.10 & +13.78
& 1.82 $\pm$ 0.71 & 2.02 $\pm$ 0.77 \\

\midrule
Unified & Unified
& +0.96 & +5.99
& +0.74 & +7.37
& +0.97 & +9.62
& +1.09 & +10.81
& +1.14 & +14.32
& 2.55 $\pm$ 1.78 & 2.94 $\pm$ 1.87 \\

\bottomrule
\end{tabular}
\vspace{-1mm}
\end{table*}

\stitle{Evaluation Metrics}
For each benchmark instance $i$, let $\pi_i(a)$ and $\pi_i'(a)$ denote the target advertisement's ranks under control and treatment, respectively, starting from 1. We define the rank gain as
\begin{equation}
\Delta r_i = \pi_i(a) - \pi_i'(a).
\end{equation}
A positive value indicates that the target advertisement moves to a higher rank. We report two metrics: \textit{$\Delta$ Rank}, the average rank gain across benchmark instances, and \textit{$\Delta$ Win}, the change in the fraction of targets ranked first between treatment and control. Both metrics are calculated separately for each dataset and averaged across three repetitions.

\subsection{Model Configurations}
\label{sec:model-config}

We use generation models for Hint Collectors and the Hint Aggregator, and ranking models for advertisement ranking. Within each generation configuration, all collectors and the aggregator use the same model. We consider two fundamental ranking approaches: embedding-based similarity scoring~\citep{grbovic2016scalable,hu2026gem,zhang2026naiad} and LLM-based listwise ranking~\citep{sun2023chatgpt,zhuang2024setwise}. Both are commonly used as components of more complex retrieval and ranking pipelines, and we evaluate them separately to examine their responses to context hints.

\stitle{Text-Embedding Ranker}
For each benchmark instance $i$, the ranker computes scores $s_i(q,b)$ and $s_i'(q,b)$ for every advertisement $b\in\mathcal{A}_q$ under control and treatment, respectively. Sorting advertisements by descending score produces rankings $\pi_i$ and $\pi_i'$. Under control, advertisements are scored using ad-card similarity:
\begin{equation*}
s_i(q,b)=\mathrm{sim}(\mathbf{q},\mathbf{c}_b).
\end{equation*}
Under treatment, context hints are incorporated into the target advertisement's score:
\begin{equation*}
s_i'(q,b)=
\begin{cases}
\lambda\,\mathrm{sim}(\mathbf{q},\mathbf{c}_a)
+(1-\lambda)\displaystyle\max_{h\in\mathcal{H}_a}
\mathrm{sim}(\mathbf{q},\mathbf{h}),
& b=a,\\[4pt]
\mathrm{sim}(\mathbf{q},\mathbf{c}_b),
& b\neq a.
\end{cases}
\end{equation*}
Here, $\mathbf{q}$, $\mathbf{c}_b$, and $\mathbf{h}$ are the embeddings of the query, ad card $\mathcal{C}_b$, and context hint $h$, respectively. $\mathrm{sim}(\cdot,\cdot)$ denotes cosine similarity, and $\lambda\in[0,1]$ controls the relative weights of ad-card and context-hint similarities. We set $\lambda=0.5$ by default and use each embedding model's default dimensionality. We also examine sensitivity to $\lambda$.

\stitle{LLM Ranker}
We invoke the LLM ranker separately under control and treatment. Under control, the ranker ranks all candidate advertisements using their ad cards alone. Under treatment, it first selects the context hint in $\mathcal{H}_a$ most relevant to the query, then ranks the candidate pool using the target's ad card and selected hint alongside the ad cards of all other advertisements. The ranker produces listwise rankings $\pi_i$ and $\pi_i'$ for the two conditions.

\stitle{Model Selection}
We use \texttt{GPT-6-\allowbreak Luna} and \texttt{Gemini-3.1-flash-\allowbreak lite} for hint generation, \texttt{OpenAI-text-\allowbreak embedding-3-\allowbreak large} and \texttt{Qwen3-\allowbreak Embedding-\allowbreak 4B} for embedding-based ranking, and \texttt{GPT-5.6-\allowbreak Luna} and \texttt{DeepSeek-\allowbreak 4.1-flash} for LLM-based ranking.
We set the temperature to 0 for LLM-based ranking and to 0.7 for hint generation with \texttt{GPT-6-\allowbreak Luna}, while using the default temperature for \texttt{Gemini-3.1-flash-\allowbreak lite}. We independently sample three repetitions for hint generation.
The two generation models and four ranking models yield eight generation--ranking combinations, which we evaluate across five datasets and eight hint generation strategies. Using models from different families also allows us to examine whether the observed ranking gains depend on the relationship between generation and ranking models.

\definecolor{rankOpenAI}{HTML}{3F6389}
\definecolor{rankQwenE}{HTML}{4F8577}
\definecolor{rankGPT}{HTML}{765F8C}
\definecolor{rankDeepSeek}{HTML}{927D55}
\definecolor{GenGPTSix}{HTML}{4F555D}
\definecolor{GenGemini}{HTML}{8A9199}

\pgfplotsset{
  rqtwobar/.style={
    scale only axis,
    width=0.41\columnwidth,
    height=0.3\columnwidth,
    xmin=-0.6, xmax=6.2,
    ymin=0,
    xtick=\empty,
    scaled y ticks=false,
    yticklabel style={/pgf/number format/fixed, /pgf/number format/precision=1,
      xshift=0.5pt},
    axis lines=box,
    axis line style={black, line width=0.5pt},
    tick align=inside,
    xtick pos=bottom, ytick pos=left,
    major tick length=2.2pt,
    xtick style={draw=none},
    tick style={black, line width=0.5pt},
    tick label style={font=\scriptsize},
    ymajorgrids,
    grid style={black!22, line width=0.3pt, dash pattern=on 0.6pt off 1.2pt},
    set layers,
    after end axis/.append code={%
      \draw[black!55, line width=0.4pt, densely dashed]
        ({axis cs:4.0,0} |- {rel axis cs:0,0}) -- ({axis cs:4.0,0} |- {rel axis cs:0,1});
      \node[font=\scriptsize, anchor=north, inner sep=2.5pt]
        at ({axis cs:1.5,0} |- {rel axis cs:0,0}) {\textbf{Ranker}};
      \node[font=\scriptsize, anchor=north, inner sep=2.5pt]
        at ({axis cs:5.1,0} |- {rel axis cs:0,0}) {\textbf{Generator}};},
    clip=false,
  },
  RQTwoBarBase/.style={ybar, bar width=0.62, bar shift=0pt, forget plot,
    line width=0.4pt,
    nodes near coords,
    nodes near coords style={font=\tiny, text=black, rotate=90, anchor=west,
      inner sep=1.5pt,
      /pgf/number format/fixed, /pgf/number format/fixed zerofill,
      /pgf/number format/precision=2},
    point meta=y},
  RQTwoBarEmb/.style={RQTwoBarBase, fill=#1, draw=#1!70!black},
  RQTwoBarLLM/.style={RQTwoBarBase, fill=#1!18, draw=#1!85!black,
    postaction={pattern=north east lines, pattern color=#1}},
  RQTwoBarGen/.style={RQTwoBarBase, fill=#1!18, draw=#1!85!black,
    postaction={pattern=crosshatch, pattern color=#1}},
}

\providecommand{\RQTwoPending}[1]{%
  \node[font=\tiny, text=black!45, rotate=90, anchor=west, inner sep=1.5pt]
    at (axis cs:#1,0) {pending};}

\begin{figure}[t]
\centering
\begin{tikzpicture}
  \begin{axis}[hide axis, scale only axis, width=1pt, height=1pt,
    xmin=0, xmax=1, ymin=0, ymax=1,
    legend columns=3,
    legend style={draw=black!70, line width=0.4pt, fill=white, font=\scriptsize,
      inner xsep=2.5pt, inner ysep=1.5pt, row sep=0.5pt,
      /tikz/every even column/.append style={column sep=0.16cm},
      /tikz/every odd column/.append style={column sep=0.04cm},
      legend cell align=left, anchor=center, at={(0.5,0.5)}},
    legend image code/.code={%
      \draw[#1, line width=0.4pt] (0cm,-0.09cm) rectangle (0.30cm,0.11cm);}]
    \addlegendimage{fill=rankOpenAI, draw=rankOpenAI!70!black}\addlegendentry{OpenAI Emb.}
    \addlegendimage{fill=rankGPT!18, draw=rankGPT!85!black,
      postaction={pattern=north east lines, pattern color=rankGPT}}\addlegendentry{GPT-5.6-Luna}
    \addlegendimage{fill=GenGPTSix!18, draw=GenGPTSix!85!black,
      postaction={pattern=crosshatch, pattern color=GenGPTSix}}\addlegendentry{GPT-6-Luna}
    \addlegendimage{fill=rankQwenE,  draw=rankQwenE!70!black}\addlegendentry{Qwen Emb.}
    \addlegendimage{fill=rankDeepSeek!18, draw=rankDeepSeek!85!black,
      postaction={pattern=north east lines, pattern color=rankDeepSeek}}\addlegendentry{DeepSeek-4.1-flash}
    \addlegendimage{fill=GenGemini!18, draw=GenGemini!85!black,
      postaction={pattern=crosshatch, pattern color=GenGemini}}\addlegendentry{Gemini-3.1-flash-lite}
  \end{axis}
\end{tikzpicture}

\vspace{0.3mm}
\begin{tikzpicture}
\begin{groupplot}[
  group style={group size=2 by 1, horizontal sep=0.72cm},
  rqtwobar,
]
\nextgroupplot[ymax=1.75, ytick distance=0.5]
  \addplot[RQTwoBarEmb=rankOpenAI]   coordinates {(0,1.0376)};
  \addplot[RQTwoBarEmb=rankQwenE]    coordinates {(1,1.3417)};
  \addplot[RQTwoBarLLM=rankGPT]      coordinates {(2,0.5581)};
  \addplot[RQTwoBarLLM=rankDeepSeek] coordinates {(3,0.7803)};
  \addplot[RQTwoBarGen=GenGPTSix]     coordinates {(4.6,0.8134)};
  \addplot[RQTwoBarGen=GenGemini]     coordinates {(5.6,1.0455)};
\nextgroupplot[ymax=14, ytick distance=4]
  \addplot[RQTwoBarEmb=rankOpenAI]   coordinates {(0,8.4689)};
  \addplot[RQTwoBarEmb=rankQwenE]    coordinates {(1,8.4676)};
  \addplot[RQTwoBarLLM=rankGPT]      coordinates {(2,8.6098)};
  \addplot[RQTwoBarLLM=rankDeepSeek] coordinates {(3,11.0121)};
  \addplot[RQTwoBarGen=GenGPTSix]     coordinates {(4.6,8.2085)};
  \addplot[RQTwoBarGen=GenGemini]     coordinates {(5.6,10.0707)};
\end{groupplot}
\node[font=\scriptsize\bfseries, anchor=south west, inner sep=0pt]
  at ([yshift=2mm]group c1r1.north west) {$\boldsymbol{\Delta}$ Rank};
\node[font=\scriptsize\bfseries, anchor=south west, inner sep=0pt]
  at ([yshift=2mm]group c2r1.north west) {$\boldsymbol{\Delta}$ Win (pp)};
\end{tikzpicture}
\vspace{-2mm}
\caption{Ranking gains from context hints across generation and ranking models. Ranker-level results are averaged over two generators, while generator-level results are averaged over four rankers. Both are averaged over eight hint strategies and five datasets.}
\label{fig:rq2-model}
\vspace{-2mm}
\end{figure}
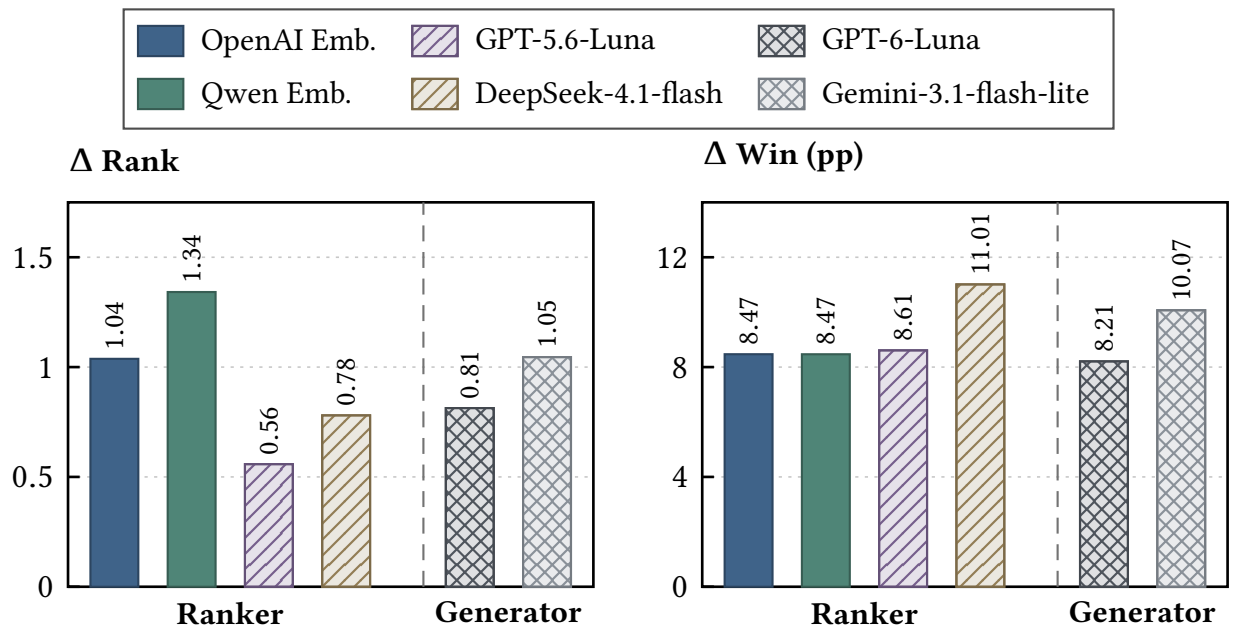

\section{Experiment Results}
\label{sec:exp}

We evaluate \ourproblem through three research questions:
\begin{itemize}[topsep=2pt,itemsep=1pt,parsep=0pt,partopsep=0pt,leftmargin=11pt]
    \item \textbf{RQ1:} Do context hints improve target-ad rankings, and how do different information strategies perform?
    \item \textbf{RQ2:} How do ranking gains and strategy preferences vary across generation and ranking models?
    \item \textbf{RQ3:} How does the hint budget affect ranking gains?
\end{itemize}

For RQ1--RQ2, we evaluate eight context hint generation strategies without a specified hint budget, allowing the aggregator to determine the number of hints. Across five datasets, two generation models, and four ranking models, this yields 320 experimental configurations. For RQ3, we require Unified to generate exactly $k$ hints, with $k\in\{3,5,20,50,100\}$, adding 200 configurations. With three repetitions, our experiments comprise \textit{520 configurations and 1,195,584 paired evaluations}. We also examine sensitivity to the embedding ranking weight $\lambda$, finding positive average gains across all evaluated settings. Full results are provided in~\trref{}. The benchmark data, generation prompts, generated context hints, and evaluation code will be made publicly available.

\definecolor{rankOpenAI}{HTML}{3F6389}
\definecolor{rankQwenE}{HTML}{4F8577}
\definecolor{rankGPT}{HTML}{765F8C}
\definecolor{rankDeepSeek}{HTML}{927D55}
\definecolor{GenGPTSix}{HTML}{4F555D}
\definecolor{GenGemini}{HTML}{8A9199}
\definecolor{DimWho}{HTML}{3F6389}
\definecolor{DimWhat}{HTML}{765F8C}
\definecolor{DimWhen}{HTML}{4F8577}

\def\yCS{8.2}\def\yCJ{7.2}
\def\yPF{5.8}\def\yPR{4.8}\def\yGC{3.8}
\def\yUS{2.4}\def\yDS{1.4}
\def\yTop{8.85}\def\yBot{0.75}

\newcommand{\rkx}[2]{#1+0.72*(8-#2)/7}
\newcommand{\rkbar}[4]{%
  \filldraw[fill=#4, draw=#4!70!black, line width=0.4pt] (axis cs:#1,{#3-0.36}) rectangle (axis cs:{\rkx{#1}{#2}},{#3+0.36});
  \ifnum#2<6
    \node[font=\tiny\bfseries, text=white, anchor=east, inner sep=1.2pt]
      at (axis cs:{\rkx{#1}{#2}},#3) {#2};
  \else
    \node[font=\tiny, text=#4!85!black, anchor=west, inner sep=1.2pt]
      at (axis cs:{\rkx{#1}{#2}},#3) {#2};
  \fi}
\newcommand{\ranklane}[8]{%
  \rkbar{#1}{#2}{\yCS}{DimWho}
  \rkbar{#1}{#3}{\yCJ}{DimWho}
  \rkbar{#1}{#4}{\yPF}{DimWhat}
  \rkbar{#1}{#5}{\yPR}{DimWhat}
  \rkbar{#1}{#6}{\yGC}{DimWhat}
  \rkbar{#1}{#7}{\yUS}{DimWhen}
  \rkbar{#1}{#8}{\yDS}{DimWhen}
  \draw[black!85, line width=0.75pt, line join=round]
    plot[smooth, tension=0.45] coordinates {
      (axis cs:#1,\yTop)
      (axis cs:{\rkx{#1}{#2}},\yCS) (axis cs:{\rkx{#1}{#3}},\yCJ)
      (axis cs:{\rkx{#1}{#4}},\yPF) (axis cs:{\rkx{#1}{#5}},\yPR)
      (axis cs:{\rkx{#1}{#6}},\yGC) (axis cs:{\rkx{#1}{#7}},\yUS)
      (axis cs:{\rkx{#1}{#8}},\yDS)
      (axis cs:#1,\yBot)};
  \draw[black!35, line width=0.4pt] (axis cs:#1,\yBot) -- (axis cs:#1,\yTop);}
\newcommand{\lanelabel}[2]{%
  \node[font=\tiny, align=center, anchor=north, inner sep=1.5pt,
        execute at begin node={\baselineskip=6.5pt}]
    at (axis cs:{#1+0.4},\yBot) {#2};}
\newcommand{\pendinglane}[1]{%
  \draw[black!30, line width=0.4pt, densely dashed] (axis cs:#1,\yBot) -- (axis cs:#1,\yTop);
  \node[font=\scriptsize\itshape, text=black!45, rotate=90]
    at (axis cs:{#1+0.4},{(\yTop+\yBot)/2}) {pending};}

\begin{figure}[t]
\centering
\newcommand{\dimkey}[1]{\tikz[baseline=-0.6ex]\filldraw[fill=#1, draw=#1!70!black, line width=0.4pt] (0,-0.09) rectangle (0.30,0.11);}
{\setlength{\fboxsep}{2.5pt}\setlength{\fboxrule}{0.4pt}%
\fcolorbox{black!70}{white}{\scriptsize
  \dimkey{DimWho}~Who \hspace{0.22cm}
  \dimkey{DimWhat}~What \hspace{0.22cm}
  \dimkey{DimWhen}~When \hspace{0.22cm}
  \tikz[baseline=-0.6ex]\draw[black!85, line width=0.75pt] (0,0) -- (0.35,0);~rank profile}}

\vspace{0.8mm}
\begin{tikzpicture}
\begin{axis}[
  scale only axis,
  width=0.88\columnwidth,
  height=5.4cm,
  xmin=-0.12, xmax=8.42,
  ymin=0.55, ymax=9.05,
  hide x axis,
  axis y line*=left,
  y axis line style={draw=none},
  xtick=\empty,
  ytick={\yCS,\yCJ,\yPF,\yPR,\yGC,\yUS,\yDS},
  yticklabels={{Customer\\Segments}, {Customer\\Jobs}, {Product\\Features},
               {Pain\\Relievers}, {Gain\\Creators}, {Usage\\Scenarios}, {Decision\\Stages}},
  yticklabel style={font=\scriptsize, align=right, anchor=east, inner sep=2pt,
    execute at begin node={\baselineskip=7.5pt}},
  ytick style={draw=none},
  clip=false,
]
\pgfplotsinvokeforeach{\yCS,\yCJ,\yPF,\yPR,\yGC,\yUS,\yDS}{%
  \draw[black!10, line width=0.3pt] (axis cs:-0.12,#1) -- (axis cs:8.42,#1);}

\ranklane{0}{3}{4}{6}{7}{5}{2}{1}
\ranklane{1}{3}{4}{6}{7}{5}{2}{1}
\ranklane{2}{1}{2}{6}{5}{4}{3}{7}
\ranklane{3}{2}{1}{7}{3}{4}{6}{5}

\ranklane{4.4}{2}{4}{5}{7}{6}{3}{1}
\ranklane{5.4}{6}{4}{2}{7}{5}{1}{3}
\ranklane{6.4}{1}{3}{2}{7}{4}{5}{6}
\ranklane{7.4}{1}{2}{4}{7}{6}{5}{3}

\lanelabel{0}{OpenAI\\Emb.}
\lanelabel{1}{Qwen\\Emb.}
\lanelabel{2}{GPT-5.6\\Luna}
\lanelabel{3}{DeepSeek\\4.1-flash}
\lanelabel{4.4}{OpenAI\\Emb.}
\lanelabel{5.4}{Qwen\\Emb.}
\lanelabel{6.4}{GPT-5.6\\Luna}
\lanelabel{7.4}{DeepSeek\\4.1-flash}

\draw[black!70, line width=0.45pt]
  (axis cs:0,9.05) -- (axis cs:0,9.25) -- (axis cs:3.8,9.25)
  node[pos=0.5, font=\scriptsize, text=black, fill=white, inner xsep=3pt, inner ysep=0.5pt]
    {\textbf{GPT-6-Luna}}
  -- (axis cs:3.8,9.05);
\draw[black!70, line width=0.45pt]
  (axis cs:4.4,9.05) -- (axis cs:4.4,9.25) -- (axis cs:8.2,9.25)
  node[pos=0.5, font=\scriptsize, text=black, fill=white, inner xsep=3pt, inner ysep=0.5pt]
    {\textbf{Gemini-3.1-flash-lite}}
  -- (axis cs:8.2,9.05);

\end{axis}
\end{tikzpicture}
\vspace{-4mm}
\caption{Relative rankings of seven dimension-specific strategies across eight generation--ranking configurations, based on average $\Delta$ Rank gains over five datasets. Longer bars indicate better relative rankings, with rank 1 being the best. Colors distinguish What, Who, and When dimensions.}
\label{fig:rq2-config-ranks}
\vspace{-2mm}
\end{figure}

\definecolor{rankOpenAI}{HTML}{3F6389}
\definecolor{rankQwenE}{HTML}{4F8577}
\definecolor{rankGPT}{HTML}{765F8C}
\definecolor{rankDeepSeek}{HTML}{927D55}
\definecolor{GenGPTSix}{HTML}{4F555D}
\definecolor{GenGemini}{HTML}{8A9199}

\pgfdeclareplotmark{starfill}{%
  \pgfpathmoveto{\pgfpointpolar{90}{1.15\pgfplotmarksize}}%
  \pgfpathlineto{\pgfpointpolar{126}{0.47\pgfplotmarksize}}%
  \pgfpathlineto{\pgfpointpolar{162}{1.15\pgfplotmarksize}}%
  \pgfpathlineto{\pgfpointpolar{198}{0.47\pgfplotmarksize}}%
  \pgfpathlineto{\pgfpointpolar{234}{1.15\pgfplotmarksize}}%
  \pgfpathlineto{\pgfpointpolar{270}{0.47\pgfplotmarksize}}%
  \pgfpathlineto{\pgfpointpolar{306}{1.15\pgfplotmarksize}}%
  \pgfpathlineto{\pgfpointpolar{342}{0.47\pgfplotmarksize}}%
  \pgfpathlineto{\pgfpointpolar{18}{1.15\pgfplotmarksize}}%
  \pgfpathlineto{\pgfpointpolar{54}{0.47\pgfplotmarksize}}%
  \pgfpathclose\pgfusepathqfillstroke}
\pgfdeclareplotmark{crossfill}{%
  \pgfpathmoveto{\pgfqpoint{0.36\pgfplotmarksize}{0.36\pgfplotmarksize}}%
  \pgfpathlineto{\pgfqpoint{0.36\pgfplotmarksize}{\pgfplotmarksize}}%
  \pgfpathlineto{\pgfqpoint{-0.36\pgfplotmarksize}{\pgfplotmarksize}}%
  \pgfpathlineto{\pgfqpoint{-0.36\pgfplotmarksize}{0.36\pgfplotmarksize}}%
  \pgfpathlineto{\pgfqpoint{-\pgfplotmarksize}{0.36\pgfplotmarksize}}%
  \pgfpathlineto{\pgfqpoint{-\pgfplotmarksize}{-0.36\pgfplotmarksize}}%
  \pgfpathlineto{\pgfqpoint{-0.36\pgfplotmarksize}{-0.36\pgfplotmarksize}}%
  \pgfpathlineto{\pgfqpoint{-0.36\pgfplotmarksize}{-\pgfplotmarksize}}%
  \pgfpathlineto{\pgfqpoint{0.36\pgfplotmarksize}{-\pgfplotmarksize}}%
  \pgfpathlineto{\pgfqpoint{0.36\pgfplotmarksize}{-0.36\pgfplotmarksize}}%
  \pgfpathlineto{\pgfqpoint{\pgfplotmarksize}{-0.36\pgfplotmarksize}}%
  \pgfpathlineto{\pgfqpoint{\pgfplotmarksize}{0.36\pgfplotmarksize}}%
  \pgfpathclose\pgfusepathqfillstroke}

\pgfplotsset{
  budgetaxis/.style={
    scale only axis,
    width=0.159\textwidth,
    height=0.1\textwidth,
    x filter/.expression={x<5 ? 3+(x-3)*7 : 12+x},
    xmin=-0.5, xmax=116,
    xtick={3,17,32,62,112},
    xticklabels={3,5,20,50,100},
    minor xtick={},
    set layers,
    after end axis/.append code={%
      \pgfonlayer{axis background}%
      \fill[black!4] ({axis cs:-0.5,0} |- {rel axis cs:0,0}) rectangle ({axis cs:17.5,0} |- {rel axis cs:0,1});%
      \endpgfonlayer},
    enlarge y limits={value=0.10},
    scaled y ticks=false,
    yticklabel style={/pgf/number format/fixed, /pgf/number format/precision=1},
    max space between ticks=16,
    axis lines=box,
    axis line style={black, line width=0.5pt},
    tick align=inside,
    xtick pos=bottom, ytick pos=left,
    major tick length=2.2pt,
    tick style={black, line width=0.5pt},
    tick label style={font=\scriptsize},
    xticklabel style={yshift=0.5pt},
    yticklabel style={xshift=0.5pt},
    grid=major,
    grid style={black!22, line width=0.3pt, dash pattern=on 0.6pt off 1.2pt},
    label style={font=\scriptsize},
    xlabel style={font=\scriptsize, yshift=2pt},
    every axis y label/.style={at={(rel axis cs:0,1)}, anchor=south west,
      rotate=0, font=\scriptsize, inner sep=0pt, yshift=2.5pt, xshift=-0.1pt},
    title style={font=\small, yshift=5pt},
    clip=false,
  },
  GenGPT/.style={solid, line width=0.8pt, mark options={solid, line width=0.3pt}},
  GenGem/.style ={densely dashed, line width=0.8pt, mark options={solid, line width=0.3pt}},
  sGptOpenAI/.style={GenGPT, color=rankOpenAI, mark=*,         mark size=1.55pt},
  sGptQwenE/.style ={GenGPT, color=rankQwenE,  mark=square*,   mark size=1.40pt},
  sGptGPT/.style   ={GenGPT, color=rankGPT,    mark=triangle*, mark size=1.95pt},
  sGptDeepSeek/.style ={GenGPT, color=rankDeepSeek,  mark=diamond*,  mark size=2.10pt},
  sGemOpenAI/.style ={GenGem,  color=rankOpenAI, mark=starfill,  mark size=1.85pt},
  sGemQwenE/.style  ={GenGem,  color=rankQwenE,  mark=crossfill, mark size=1.75pt},
  sGemGPT/.style    ={GenGem,  color=rankGPT,    mark=triangle*, mark size=1.95pt,
                     mark options={solid, line width=0.3pt, rotate=180}},
  sGemDeepSeek/.style  ={GenGem,  color=rankDeepSeek,  mark=crossfill, mark size=1.75pt,
                     mark options={solid, line width=0.3pt, rotate=45}},
  Unlim/.style={only marks, forget plot, mark=*, mark size=2.0pt,
                mark options={solid, fill=white, line width=0.8pt}},
  ZeroLine/.style={forget plot, black!55, line width=0.4pt},
}

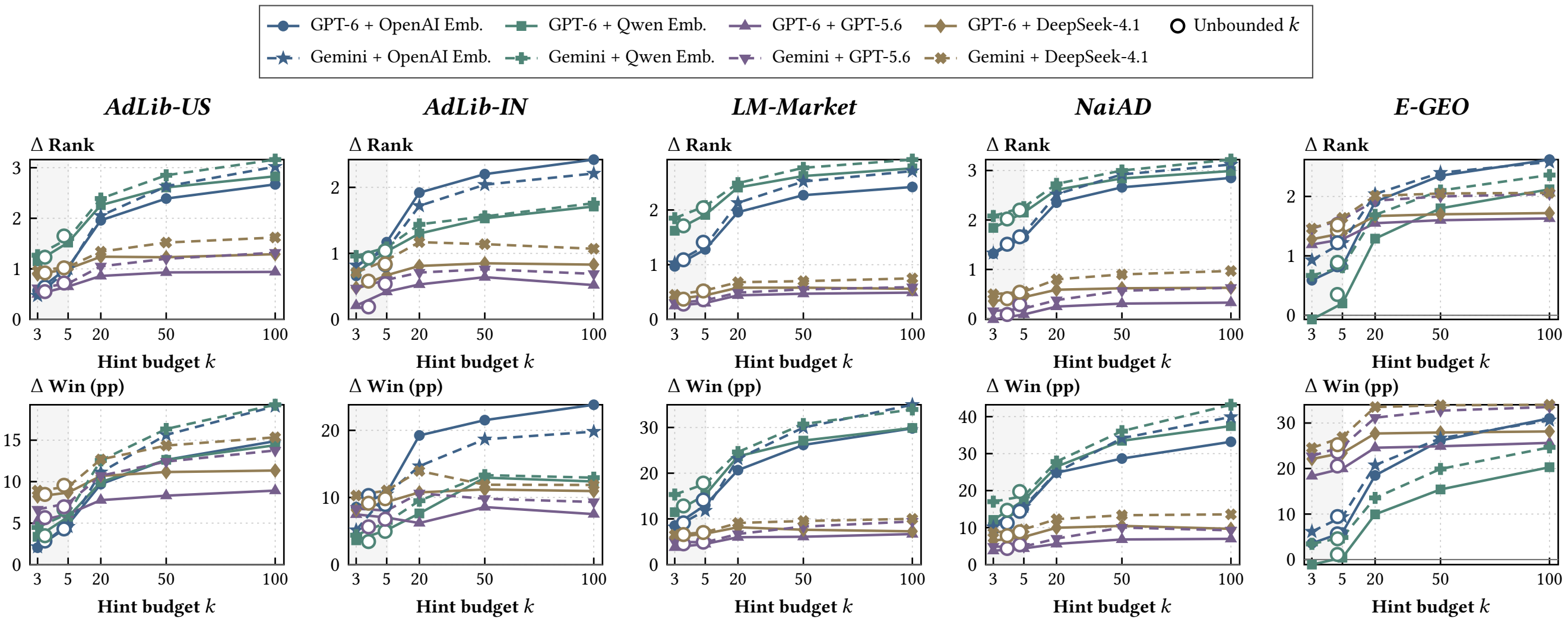
\begin{figure*}[t]
\centering
\begin{tikzpicture}
  \begin{axis}[hide axis, scale only axis, width=1pt, height=1pt,
    xmin=0, xmax=1, ymin=0, ymax=1,
    legend columns=5,
    legend style={draw=black!70, line width=0.4pt, fill=white, font=\scriptsize,
      inner xsep=2pt, inner ysep=1.5pt, row sep=0.5pt,
      /tikz/every even column/.append style={column sep=0.07cm},
      /tikz/every odd column/.append style={column sep=0.04cm},
      legend cell align=left, anchor=center, at={(0.5,0.5)}},
    legend image code/.code={\draw[mark repeat=2, mark phase=2, #1]
      plot coordinates {(0cm,0cm) (0.18cm,0cm) (0.36cm,0cm)};}]
    \addlegendimage{sGptOpenAI}\addlegendentry{GPT-6 + OpenAI Emb.}
    \addlegendimage{sGptQwenE}\addlegendentry{GPT-6 + Qwen Emb.}
    \addlegendimage{sGptGPT}\addlegendentry{GPT-6 + GPT-5.6}
    \addlegendimage{sGptDeepSeek}\addlegendentry{GPT-6 + DeepSeek-4.1}
    \addlegendimage{black, only marks, mark=*, mark size=2.0pt,
      mark options={fill=white, line width=0.8pt}}\addlegendentry{Unbounded $k$}
    \addlegendimage{sGemOpenAI}\addlegendentry{Gemini + OpenAI Emb.}
    \addlegendimage{sGemQwenE}\addlegendentry{Gemini + Qwen Emb.}
    \addlegendimage{sGemGPT}\addlegendentry{Gemini + GPT-5.6}
    \addlegendimage{sGemDeepSeek}\addlegendentry{Gemini + DeepSeek-4.1}
    \addlegendimage{empty legend}\addlegendentry{}
  \end{axis}
\end{tikzpicture}

\vspace{0.6mm}
\begin{tikzpicture}
\begin{groupplot}[
  group style={group size=5 by 2, horizontal sep=0.72cm, vertical sep=0.95cm},
  budgetaxis,
]
\nextgroupplot[title=\textbf{\adlib}, ylabel=\textbf{$\Delta$ Rank}, xlabel=\textbf{Hint budget $k$}]
\vspace{-4mm}
\addplot[ZeroLine] coordinates {(2.5,0) (104,0)};
\addplot[sGptOpenAI] coordinates {(3,0.57) (5,1.03) (20,1.96) (50,2.39) (100,2.67)};
\addplot[sGptQwenE] coordinates {(3,1.15) (5,1.52) (20,2.26) (50,2.61) (100,2.83)};
\addplot[sGptGPT] coordinates {(3,0.50) (5,0.65) (20,0.86) (50,0.93) (100,0.94)};
\addplot[sGptDeepSeek] coordinates {(3,0.89) (5,1.00) (20,1.24) (50,1.23) (100,1.29)};
\addplot[sGemOpenAI] coordinates {(3,0.47) (5,1.00) (20,2.05) (50,2.63) (100,3.02)};
\addplot[sGemQwenE] coordinates {(3,1.27) (5,1.60) (20,2.39) (50,2.85) (100,3.16)};
\addplot[sGemGPT] coordinates {(3,0.62) (5,0.71) (20,1.05) (50,1.20) (100,1.32)};
\addplot[sGemDeepSeek] coordinates {(3,0.96) (5,1.04) (20,1.34) (50,1.52) (100,1.62)};
\addplot[Unlim, rankOpenAI] coordinates {(3.48,0.64)};
\addplot[Unlim, rankQwenE] coordinates {(3.48,1.23)};
\addplot[Unlim, rankGPT] coordinates {(3.48,0.55)};
\addplot[Unlim, rankDeepSeek] coordinates {(3.48,0.92)};
\addplot[Unlim, rankOpenAI] coordinates {(4.73,0.94)};
\addplot[Unlim, rankQwenE] coordinates {(4.73,1.65)};
\addplot[Unlim, rankGPT] coordinates {(4.73,0.72)};
\addplot[Unlim, rankDeepSeek] coordinates {(4.73,1.02)};

\nextgroupplot[title=\textbf{\adindia}, ylabel=\textbf{$\Delta$ Rank}, xlabel=\textbf{Hint budget $k$}]
\vspace{-4mm}
\addplot[ZeroLine] coordinates {(2.5,0) (104,0)};
\addplot[sGptOpenAI] coordinates {(3,0.66) (5,1.17) (20,1.92) (50,2.20) (100,2.42)};
\addplot[sGptQwenE] coordinates {(3,0.71) (5,1.03) (20,1.30) (50,1.53) (100,1.71)};
\addplot[sGptGPT] coordinates {(3,0.21) (5,0.42) (20,0.53) (50,0.64) (100,0.52)};
\addplot[sGptDeepSeek] coordinates {(3,0.48) (5,0.67) (20,0.81) (50,0.85) (100,0.83)};
\addplot[sGemOpenAI] coordinates {(3,0.82) (5,1.06) (20,1.72) (50,2.04) (100,2.21)};
\addplot[sGemQwenE] coordinates {(3,0.96) (5,1.09) (20,1.44) (50,1.56) (100,1.76)};
\addplot[sGemGPT] coordinates {(3,0.47) (5,0.60) (20,0.71) (50,0.76) (100,0.69)};
\addplot[sGemDeepSeek] coordinates {(3,0.71) (5,0.90) (20,1.17) (50,1.14) (100,1.07)};
\addplot[Unlim, rankOpenAI] coordinates {(3.81,0.93)};
\addplot[Unlim, rankQwenE] coordinates {(3.81,0.93)};
\addplot[Unlim, rankGPT] coordinates {(3.81,0.19)};
\addplot[Unlim, rankDeepSeek] coordinates {(3.81,0.58)};
\addplot[Unlim, rankOpenAI] coordinates {(4.90,0.83)};
\addplot[Unlim, rankQwenE] coordinates {(4.90,1.04)};
\addplot[Unlim, rankGPT] coordinates {(4.90,0.54)};
\addplot[Unlim, rankDeepSeek] coordinates {(4.90,0.84)};

\nextgroupplot[title=\textbf{\lmmarket}, ylabel=\textbf{$\Delta$ Rank}, xlabel=\textbf{Hint budget $k$}]
\vspace{-4mm}
\addplot[ZeroLine] coordinates {(2.5,0) (104,0)};
\addplot[sGptOpenAI] coordinates {(3,0.97) (5,1.28) (20,1.96) (50,2.27) (100,2.42)};
\addplot[sGptQwenE] coordinates {(3,1.62) (5,1.91) (20,2.41) (50,2.62) (100,2.76)};
\addplot[sGptGPT] coordinates {(3,0.25) (5,0.30) (20,0.44) (50,0.47) (100,0.49)};
\addplot[sGptDeepSeek] coordinates {(3,0.37) (5,0.44) (20,0.58) (50,0.58) (100,0.56)};
\addplot[sGemOpenAI] coordinates {(3,1.03) (5,1.34) (20,2.13) (50,2.52) (100,2.71)};
\addplot[sGemQwenE] coordinates {(3,1.85) (5,2.07) (20,2.49) (50,2.77) (100,2.92)};
\addplot[sGemGPT] coordinates {(3,0.30) (5,0.35) (20,0.49) (50,0.55) (100,0.59)};
\addplot[sGemDeepSeek] coordinates {(3,0.45) (5,0.53) (20,0.68) (50,0.70) (100,0.75)};
\addplot[Unlim, rankOpenAI] coordinates {(3.57,1.09)};
\addplot[Unlim, rankQwenE] coordinates {(3.57,1.71)};
\addplot[Unlim, rankGPT] coordinates {(3.57,0.28)};
\addplot[Unlim, rankDeepSeek] coordinates {(3.57,0.37)};
\addplot[Unlim, rankOpenAI] coordinates {(4.87,1.41)};
\addplot[Unlim, rankQwenE] coordinates {(4.87,2.04)};
\addplot[Unlim, rankGPT] coordinates {(4.87,0.36)};
\addplot[Unlim, rankDeepSeek] coordinates {(4.87,0.52)};

\nextgroupplot[title=\textbf{\naiad}, ylabel=\textbf{$\Delta$ Rank}, xlabel=\textbf{Hint budget $k$}]
\vspace{-4mm}
\addplot[ZeroLine] coordinates {(2.5,0) (104,0)};
\addplot[sGptOpenAI] coordinates {(3,1.32) (5,1.65) (20,2.35) (50,2.66) (100,2.85)};
\addplot[sGptQwenE] coordinates {(3,1.84) (5,2.15) (20,2.61) (50,2.84) (100,2.99)};
\addplot[sGptGPT] coordinates {(3,-0.01) (5,0.09) (20,0.25) (50,0.31) (100,0.33)};
\addplot[sGptDeepSeek] coordinates {(3,0.36) (5,0.44) (20,0.59) (50,0.62) (100,0.63)};
\addplot[sGemOpenAI] coordinates {(3,1.33) (5,1.72) (20,2.52) (50,2.92) (100,3.12)};
\addplot[sGemQwenE] coordinates {(3,2.08) (5,2.26) (20,2.73) (50,3.00) (100,3.22)};
\addplot[sGemGPT] coordinates {(3,0.16) (5,0.21) (20,0.38) (50,0.57) (100,0.63)};
\addplot[sGemDeepSeek] coordinates {(3,0.50) (5,0.55) (20,0.80) (50,0.90) (100,0.97)};
\addplot[Unlim, rankOpenAI] coordinates {(3.91,1.51)};
\addplot[Unlim, rankQwenE] coordinates {(3.91,2.02)};
\addplot[Unlim, rankGPT] coordinates {(3.91,0.09)};
\addplot[Unlim, rankDeepSeek] coordinates {(3.91,0.41)};
\addplot[Unlim, rankOpenAI] coordinates {(4.73,1.66)};
\addplot[Unlim, rankQwenE] coordinates {(4.73,2.20)};
\addplot[Unlim, rankGPT] coordinates {(4.73,0.29)};
\addplot[Unlim, rankDeepSeek] coordinates {(4.73,0.54)};

\nextgroupplot[title=\textbf{\egeo}, ylabel=\textbf{$\Delta$ Rank}, xlabel=\textbf{Hint budget $k$}]
\vspace{-4mm}
\addplot[ZeroLine] coordinates {(2.5,0) (104,0)};
\addplot[sGptOpenAI] coordinates {(3,0.59) (5,0.83) (20,1.91) (50,2.35) (100,2.62)};
\addplot[sGptQwenE] coordinates {(3,-0.07) (5,0.20) (20,1.29) (50,1.80) (100,2.12)};
\addplot[sGptGPT] coordinates {(3,1.19) (5,1.28) (20,1.55) (50,1.60) (100,1.63)};
\addplot[sGptDeepSeek] coordinates {(3,1.28) (5,1.38) (20,1.67) (50,1.70) (100,1.72)};
\addplot[sGemOpenAI] coordinates {(3,0.93) (5,1.22) (20,2.04) (50,2.40) (100,2.58)};
\addplot[sGemQwenE] coordinates {(3,0.67) (5,0.81) (20,1.69) (50,2.10) (100,2.36)};
\addplot[sGemGPT] coordinates {(3,1.46) (5,1.61) (20,1.93) (50,2.00) (100,2.04)};
\addplot[sGemDeepSeek] coordinates {(3,1.46) (5,1.64) (20,2.01) (50,2.05) (100,2.06)};
\addplot[Unlim, rankOpenAI] coordinates {(4.67,0.85)};
\addplot[Unlim, rankQwenE] coordinates {(4.67,0.35)};
\addplot[Unlim, rankGPT] coordinates {(4.67,1.32)};
\addplot[Unlim, rankDeepSeek] coordinates {(4.67,1.43)};
\addplot[Unlim, rankOpenAI] coordinates {(4.68,1.22)};
\addplot[Unlim, rankQwenE] coordinates {(4.68,0.89)};
\addplot[Unlim, rankGPT] coordinates {(4.68,1.53)};
\addplot[Unlim, rankDeepSeek] coordinates {(4.68,1.52)};

\nextgroupplot[ylabel=\textbf{$\Delta$ Win (pp)}, xlabel=\textbf{Hint budget $k$}]
\vspace{-4mm}
\addplot[ZeroLine] coordinates {(2.5,0) (104,0)};
\addplot[sGptOpenAI] coordinates {(3,2.04) (5,4.85) (20,9.70) (50,12.63) (100,14.88)};
\addplot[sGptQwenE] coordinates {(3,3.35) (5,5.76) (20,9.94) (50,12.63) (100,14.39)};
\addplot[sGptGPT] coordinates {(3,5.26) (5,6.08) (20,7.77) (50,8.31) (100,8.92)};
\addplot[sGptDeepSeek] coordinates {(3,8.21) (5,8.68) (20,10.69) (50,11.16) (100,11.34)};
\addplot[sGemOpenAI] coordinates {(3,2.13) (5,4.57) (20,11.13) (50,15.64) (100,19.09)};
\addplot[sGemQwenE] coordinates {(3,4.48) (5,6.16) (20,12.56) (50,16.35) (100,19.27)};
\addplot[sGemGPT] coordinates {(3,6.64) (5,7.41) (20,10.71) (50,12.42) (100,13.76)};
\addplot[sGemDeepSeek] coordinates {(3,8.92) (5,9.38) (20,12.68) (50,14.35) (100,15.35)};
\addplot[Unlim, rankOpenAI] coordinates {(3.48,2.84)};
\addplot[Unlim, rankQwenE] coordinates {(3.48,3.48)};
\addplot[Unlim, rankGPT] coordinates {(3.48,5.66)};
\addplot[Unlim, rankDeepSeek] coordinates {(3.48,8.46)};
\addplot[Unlim, rankOpenAI] coordinates {(4.73,4.30)};
\addplot[Unlim, rankQwenE] coordinates {(4.73,6.65)};
\addplot[Unlim, rankGPT] coordinates {(4.73,6.97)};
\addplot[Unlim, rankDeepSeek] coordinates {(4.73,9.56)};

\nextgroupplot[ylabel=\textbf{$\Delta$ Win (pp)}, xlabel=\textbf{Hint budget $k$}]
\vspace{-4mm}
\addplot[ZeroLine] coordinates {(2.5,0) (104,0)};
\addplot[sGptOpenAI] coordinates {(3,8.57) (5,10.67) (20,19.24) (50,21.52) (100,23.81)};
\addplot[sGptQwenE] coordinates {(3,3.62) (5,5.14) (20,7.62) (50,12.95) (100,12.38)};
\addplot[sGptGPT] coordinates {(3,7.48) (5,7.00) (20,6.19) (50,8.57) (100,7.52)};
\addplot[sGptDeepSeek] coordinates {(3,8.14) (5,9.33) (20,10.76) (50,11.19) (100,10.95)};
\addplot[sGemOpenAI] coordinates {(3,5.14) (5,10.48) (20,14.67) (50,18.67) (100,19.81)};
\addplot[sGemQwenE] coordinates {(3,4.38) (5,6.29) (20,9.52) (50,13.33) (100,12.95)};
\addplot[sGemGPT] coordinates {(3,8.33) (5,8.05) (20,10.62) (50,9.81) (100,9.33)};
\addplot[sGemDeepSeek] coordinates {(3,10.29) (5,11.05) (20,13.95) (50,11.90) (100,11.86)};
\addplot[Unlim, rankOpenAI] coordinates {(3.81,10.29)};
\addplot[Unlim, rankQwenE] coordinates {(3.81,3.43)};
\addplot[Unlim, rankGPT] coordinates {(3.81,5.62)};
\addplot[Unlim, rankDeepSeek] coordinates {(3.81,9.14)};
\addplot[Unlim, rankOpenAI] coordinates {(4.90,8.95)};
\addplot[Unlim, rankQwenE] coordinates {(4.90,4.95)};
\addplot[Unlim, rankGPT] coordinates {(4.90,6.76)};
\addplot[Unlim, rankDeepSeek] coordinates {(4.90,9.81)};

\nextgroupplot[ylabel=\textbf{$\Delta$ Win (pp)}, xlabel=\textbf{Hint budget $k$}]
\vspace{-4mm}
\addplot[ZeroLine] coordinates {(2.5,0) (104,0)};
\addplot[sGptOpenAI] coordinates {(3,8.53) (5,13.01) (20,20.67) (50,26.18) (100,29.80)};
\addplot[sGptQwenE] coordinates {(3,11.46) (5,15.25) (20,23.69) (50,27.13) (100,29.89)};
\addplot[sGptGPT] coordinates {(3,3.77) (5,4.52) (20,5.99) (50,6.09) (100,6.68)};
\addplot[sGptDeepSeek] coordinates {(3,5.99) (5,6.76) (20,8.16) (50,7.62) (100,7.28)};
\addplot[sGemOpenAI] coordinates {(3,8.18) (5,11.80) (20,23.17) (50,29.97) (100,34.97)};
\addplot[sGemQwenE] coordinates {(3,15.33) (5,17.66) (20,24.63) (50,30.75) (100,33.94)};
\addplot[sGemGPT] coordinates {(3,4.97) (5,4.89) (20,6.70) (50,8.31) (100,9.43)};
\addplot[sGemDeepSeek] coordinates {(3,6.52) (5,7.08) (20,9.11) (50,9.54) (100,10.01)};
\addplot[Unlim, rankOpenAI] coordinates {(3.57,9.13)};
\addplot[Unlim, rankQwenE] coordinates {(3.57,12.83)};
\addplot[Unlim, rankGPT] coordinates {(3.57,4.54)};
\addplot[Unlim, rankDeepSeek] coordinates {(3.57,6.57)};
\addplot[Unlim, rankOpenAI] coordinates {(4.87,14.13)};
\addplot[Unlim, rankQwenE] coordinates {(4.87,17.83)};
\addplot[Unlim, rankGPT] coordinates {(4.87,4.91)};
\addplot[Unlim, rankDeepSeek] coordinates {(4.87,7.02)};

\nextgroupplot[ylabel=\textbf{$\Delta$ Win (pp)}, xlabel=\textbf{Hint budget $k$}]
\vspace{-4mm}
\addplot[ZeroLine] coordinates {(2.5,0) (104,0)};
\addplot[sGptOpenAI] coordinates {(3,10.97) (5,15.09) (20,24.83) (50,28.67) (100,33.20)};
\addplot[sGptQwenE] coordinates {(3,12.07) (5,17.42) (20,26.47) (50,33.47) (100,37.45)};
\addplot[sGptGPT] coordinates {(3,3.74) (5,4.36) (20,5.59) (50,6.79) (100,6.96)};
\addplot[sGptDeepSeek] coordinates {(3,6.48) (5,7.48) (20,9.95) (50,10.46) (100,9.67)};
\addplot[sGemOpenAI] coordinates {(3,10.15) (5,15.64) (20,24.97) (50,34.16) (100,39.92)};
\addplot[sGemQwenE] coordinates {(3,17.01) (5,18.52) (20,27.85) (50,36.08) (100,43.21)};
\addplot[sGemGPT] coordinates {(3,4.84) (5,4.87) (20,7.03) (50,10.01) (100,9.26)};
\addplot[sGemDeepSeek] coordinates {(3,8.64) (5,9.53) (20,12.28) (50,13.34) (100,13.58)};
\addplot[Unlim, rankOpenAI] coordinates {(3.91,11.25)};
\addplot[Unlim, rankQwenE] coordinates {(3.91,14.68)};
\addplot[Unlim, rankGPT] coordinates {(3.91,4.36)};
\addplot[Unlim, rankDeepSeek] coordinates {(3.91,7.82)};
\addplot[Unlim, rankOpenAI] coordinates {(4.73,14.40)};
\addplot[Unlim, rankQwenE] coordinates {(4.73,19.75)};
\addplot[Unlim, rankGPT] coordinates {(4.73,5.35)};
\addplot[Unlim, rankDeepSeek] coordinates {(4.73,8.85)};

\nextgroupplot[ylabel=\textbf{$\Delta$ Win (pp)}, xlabel=\textbf{Hint budget $k$}]
\vspace{-4mm}
\addplot[ZeroLine] coordinates {(2.5,0) (104,0)};
\addplot[sGptOpenAI] coordinates {(3,3.57) (5,5.94) (20,18.49) (50,26.15) (100,31.02)};
\addplot[sGptQwenE] coordinates {(3,-1.10) (5,0.41) (20,9.94) (50,15.41) (100,20.29)};
\addplot[sGptGPT] coordinates {(3,18.36) (5,19.98) (20,24.57) (50,24.89) (100,25.64)};
\addplot[sGptDeepSeek] coordinates {(3,22.10) (5,23.41) (20,27.70) (50,27.92) (100,28.15)};
\addplot[sGemOpenAI] coordinates {(3,6.15) (5,9.39) (20,20.75) (50,26.66) (100,30.59)};
\addplot[sGemQwenE] coordinates {(3,3.45) (5,4.86) (20,13.55) (50,19.91) (100,24.65)};
\addplot[sGemGPT] coordinates {(3,22.74) (5,24.84) (20,31.17) (50,32.72) (100,33.52)};
\addplot[sGemDeepSeek] coordinates {(3,24.51) (5,26.91) (20,33.51) (50,33.91) (100,34.01)};
\addplot[Unlim, rankOpenAI] coordinates {(4.67,5.65)};
\addplot[Unlim, rankQwenE] coordinates {(4.67,1.12)};
\addplot[Unlim, rankGPT] coordinates {(4.67,20.56)};
\addplot[Unlim, rankDeepSeek] coordinates {(4.67,24.38)};
\addplot[Unlim, rankOpenAI] coordinates {(4.68,9.45)};
\addplot[Unlim, rankQwenE] coordinates {(4.68,4.57)};
\addplot[Unlim, rankGPT] coordinates {(4.68,23.63)};
\addplot[Unlim, rankDeepSeek] coordinates {(4.68,25.24)};
\end{groupplot}
\end{tikzpicture}
\vspace{-2mm}
\caption{Effects of hint budget $k$ on Unified across five datasets. Rows show $\Delta$ Rank and $\Delta$ Win, with curves representing eight generation--ranking configurations. Hollow markers indicate unconstrained generation at the average number of generated hints. The $k=3$--$5$ interval is expanded for readability.}
\label{fig:rq3-budget}
\vspace{-2mm}
\end{figure*}

\subsection{RQ1: Hint Effectiveness and Strategies}

We first examine whether context hints improve advertisement rankings and how their effectiveness varies across information strategies. Table~\ref{tab:rq1-main} compares eight strategies across five datasets, averaging gains over eight generation--ranking model combinations. We also rank the strategies within each of the 40 dataset--generator--ranker settings and report their average \textit{relative ranking} and standard deviation. The three dimensions are compared based on their mean strategy gains within each setting. Lower relative rankings indicate better performance.

\stitle{Context hints improve target-ad rankings}
As shown in Table~\ref{tab:rq1-main}, all eight strategies yield positive average gains across the five datasets, with $\Delta$ Rank improvements of 0.56--1.16 positions and $\Delta$ Win improvements of 5.42--14.40 percentage points. Without a specified hint budget, the strategies generate 3.8--4.3 hints per advertisement on average, each containing 90--96 characters, including spaces. Each hint is shorter on average than the combined 150-character limit for ad-card titles and descriptions. Embedding rankers also achieve positive gains when ad-card and context-hint similarities receive equal weight ($\lambda=0.5$). These results show that concise context hints can improve advertisement rankings under the evaluated offline rankers.

\stitle{Information dimensions show similar gains, with some significant differences}
Among the three dimensions, Who achieves the best average relative ranking, followed by When and What, under both metrics. Pairwise comparisons show that Who and When significantly outperform What, but their absolute advantages are small, ranging from 0.061 to 0.062 positions in $\Delta$ Rank and 0.42 to 0.52 percentage points in $\Delta$ Win (Holm-adjusted $p<0.001$). Who and When do not differ significantly under either metric (Holm-adjusted $p\geq0.59$). Among individual strategies, Customer Segments achieves the best relative ranking under both metrics, followed by Customer Jobs. Overall, the evaluated information strategies achieve broadly comparable average gains.

\stitle{Consistent gains across advertising settings}
All five datasets show positive ranking gains, including \egeo, which focuses on explicit product recommendation queries. In particular, \egeo achieves $\Delta$ Win gains of 13.16--14.40 percentage points across strategies, compared with 5.42--6.01 on \adlib. Who and When also achieve higher average $\Delta$ Rank gains than What on every dataset, although the differences remain small. These results show that the observed gains extend beyond conversations with implicit commercial needs to explicit product recommendation settings.

\stitle{Unified does not consistently outperform individual strategies}
Unified achieves the best overall relative ranking under both $\Delta$ Rank (2.55) and $\Delta$ Win (2.94), but its advantages over individual strategies are limited. For $\Delta$ Rank, Unified matches the best individual strategy to two decimal places on three datasets but falls behind by 0.06 positions on \adindia and 0.02 positions on \egeo. For $\Delta$ Win, Unified leads only on \lmmarket, by 0.17 percentage points, while individual strategies perform better on the other four datasets. These results show that combining information from all seven dimensions does not necessarily yield larger gains, motivating further research on effective hint selection and aggregation.

\subsection{RQ2: Generation and Ranking Models}

We examine how generation and ranking models affect ranking gains and the relative performance of context hint generation strategies.

\stitle{Both generation and ranking models affect ranking gains}
Figure~\ref{fig:rq2-model} shows positive average $\Delta$ Rank and $\Delta$ Win gains across all four rankers and both generators. Gemini-generated hints achieve larger average gains than GPT-6-generated hints. Within ranking approaches, Qwen embeddings achieve larger average $\Delta$ Rank gains than OpenAI embeddings, while DeepSeek achieves larger gains than GPT-5.6 among LLM rankers. These results show that both generation and ranking model choices affect the measured gains from context hints.

\stitle{Embedding rankers gain more in rank, while LLM rankers gain more in top-1 placement}
Embedding rankers achieve larger average $\Delta$ Rank gains than LLM rankers (1.19 vs. 0.67 positions), whereas LLM rankers achieve larger $\Delta$ Win gains (9.81 vs. 8.47 percentage points). This difference may reflect how the two approaches incorporate context hints. Embedding rankers combine ad-card similarity with the maximum query--hint similarity, while LLM rankers select a query-relevant hint before listwise ranking. These mechanisms may produce different ranking responses to the same hints. The contrasting results also show that larger average rank gains do not necessarily yield larger top-1 gains.

\stitle{Strategy differences become more pronounced across model configurations}
Although information strategies achieve similar average gains in RQ1, their differences become more pronounced under some generation--ranking configurations. For example, on \adindia, Decision Stages achieves a $\Delta$ Rank gain of 1.03 positions with GPT-6 and OpenAI embeddings, compared with 0.46 for Customer Jobs. At the dimension level, When achieves an average gain of 0.83 positions, compared with 0.59 for Who. Figure~\ref{fig:rq2-config-ranks} further shows that When is favored by embedding-based rankers, whereas Who is favored by LLM-based rankers under $\Delta$ Rank. These preferences are less consistent under $\Delta$ Win and vary across generation models. Together, these findings show that averaging across models can obscure meaningful differences among information strategies, motivating context hint optimization tailored to different ranking configurations.

\subsection{RQ3: Hint Budget}

We examine how the hint budget affects target-ad ranking gains. Figure~\ref{fig:rq3-budget} presents the results across five datasets and eight generation--ranking configurations. We also include unconstrained generation, with hollow markers indicating the average number of generated hints and corresponding gains.

\stitle{Larger hint budgets yield diminishing ranking gains}
Unconstrained Unified achieves ranking gains comparable to those with exactly three or five hints. Increasing $k$ generally improves both $\Delta$ Rank and $\Delta$ Win across datasets and model configurations. For example, on \adlib, Gemini with Qwen embeddings achieves $\Delta$ Rank gains of 1.27 at $k=3$, 2.39 at $k=20$, and 3.16 at $k=100$. However, the marginal gains diminish as $k$ increases, with the extent varying across datasets and rankers. These results show that increasing the hint budget can improve rankings beyond unconstrained generation, while additional hints yield progressively smaller benefits.

\stitle{Embedding rankers benefit more from larger hint budgets}
Across five datasets and both generators, embedding rankers generally achieve larger improvements from $k=3$ to $k=100$ than LLM rankers under both metrics. For example, on \lmmarket, increasing $k$ from 3 to 100 raises $\Delta$ Rank by 1.07 positions for Gemini with Qwen embeddings, compared with 0.29 positions for Gemini with GPT-5.6. This difference may reflect how the rankers use context hints. Embedding rankers use the maximum query--hint similarity across all generated hints, while LLM rankers select one hint for listwise ranking. For embedding rankers, considering more hints also increases the opportunities to obtain a higher maximum query--hint similarity. Nevertheless, embedding rankers do not always achieve larger absolute gains, particularly on \egeo under $\Delta$ Win. These findings suggest that effective hint budgets depend on the ranking approach, motivating further research on budget-aware hint generation and selection.

\section{Conclusion}

We formulate Context Hint Optimization as an ad-side ranking optimization problem in LLM advertising and introduce \ourbench to evaluate context hint generation strategies under fixed offline rankers. Our experiments show that context hints improve advertisement rankings. While individual strategies achieve similar average gains, their differences become more pronounced under specific generation--ranking configurations, and combining multiple dimensions does not consistently improve performance. Larger hint budgets yield diminishing gains, with different responses across ranking approaches. These findings motivate model-specific hint selection and budget-aware generation and aggregation. Our evaluation measures ranking changes rather than independently verified ad relevance or outcomes when competing advertisements also use context hints. Future work includes relevance-aware and competitive evaluations, real-world advertising experiments, and more effective context hint optimization methods.

\bibliographystyle{ACM-Reference-Format}
\bibliography{refs}

\appendix

\section{Appendix}
\label{appendix}

\subsection{Additional Data Preprocessing Details}
\label{app:data}

We provide additional details on query filtering, candidate pool construction, and ad card and corpus preparation for the datasets introduced in Section~\ref{sec:data}.

\stitle{Preprocessing of \adlib and \adindia}
For \adlib~\citep{lurie2026beginning}, we aggregate advertisements by query, remove duplicate observations, and retain queries associated with at least two distinct advertiser brands to support competitive ranking.
For \adindia~\citep{viren2026chatgptads,trucommerce2026chatgptads}, we exclude queries with insufficient context, such as unresolved pronouns, and advertisements with incomplete or duplicate cards. We group prompts expressing the same need but differing in their closing instructions. For example, requests to compare or recommend three mesh Wi-Fi systems in Pune belong to the same group, with 1.64 queries per group on average. Each query's candidate pool initially includes all advertisements observed within its group. If the pool contains fewer than 10 distinct brands, we supplement it with semantically similar advertisements from \adindia, following \citet{hu2026gem}.
For both datasets, we preserve the original ad cards. To construct the ad corpus $\mathcal{D}_a$, we provide an LLM with the ad card $\mathcal{C}_a$, advertiser name, and category. The LLM uses web search to identify the advertiser's official website or a retailer page for the advertised product and extracts factual statements with source URLs.

\stitle{Preprocessing of \lmmarket and \naiad}
For both datasets, two LLM judges independently assess query suitability for advertising. We retain queries accepted by both judges and manually verify the results. Candidate advertisements are retrieved from the shared library described in Section~\ref{sec:data}, with distinct-brand constraints to avoid duplicate advertisers within each pool. We construct ad cards $\mathcal{C}_a$ from advertiser names and descriptions and obtain ad corpora $\mathcal{D}_a$ using the same LLM-assisted web-search procedure.

\stitle{Preprocessing of \egeo}
We apply the same two-judge screening and manual verification to the held-out test queries from \egeo~\citep{bagga2025egeo}, additionally requiring candidate products from at least two distinct brands. Unlike the other datasets, \egeo provides structured product metadata. We construct $\mathcal{D}_a$ from the original product title, description, bullet-point features, and details, without web search. To construct $\mathcal{C}_a$, an LLM extracts the brand and product type or model from the title and one or two key selling points from the description and features, subject to the 50-character title and 100-character description limits.

\subsection{Additional Experimental Results}
\label{app:additional-results}

\stitle{Full Results for RQ1}
Table~\ref{tab:appendix-full} reports the results of all eight context hint generation strategies across five datasets and eight generation--ranking configurations, complementing the aggregated results in Table~\ref{tab:rq1-main}. We assess statistical significance using paired $t$-tests with standard errors clustered by target advertisement and Holm correction across the 320 comparisons for each metric ($p<0.05$).
Among the 320 configurations, 292 (91.3\%) show statistically significant positive $\Delta$ Rank gains, and 256 (80.0\%) show significant positive $\Delta$ Win gains, with no significantly negative effects. The strength of statistical evidence varies across datasets and models. On \adindia, many $\Delta$ Win gains are not significant despite positive averages, whereas all gains on \lmmarket are significant under both metrics. On \naiad, several strategies paired with GPT-5.6 yield non-significant $\Delta$ Rank gains despite positive $\Delta$ Win gains. These results show that the aggregate improvements in RQ1 are supported by most, but not all, individual configurations.

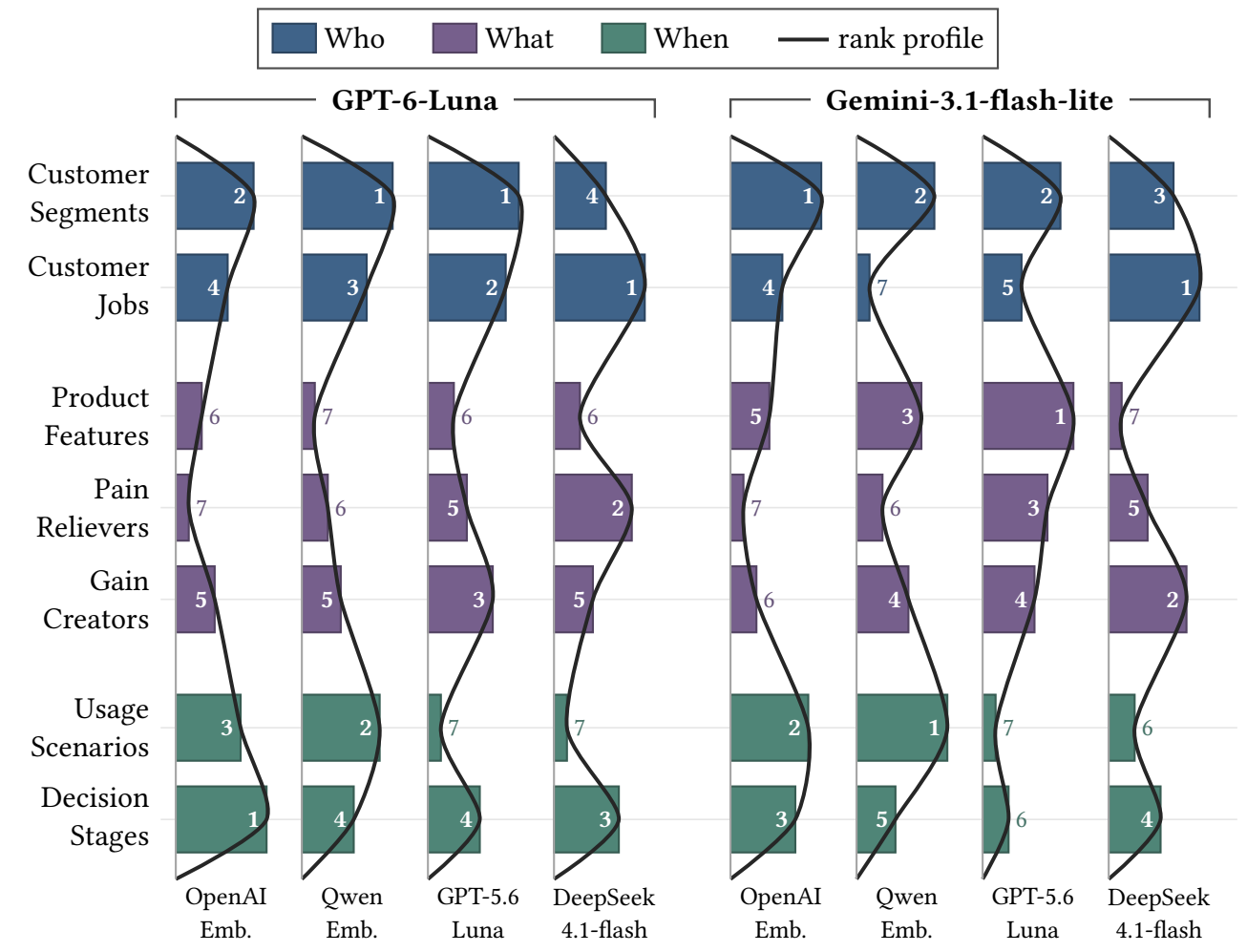
\begin{figure}[t]
\centering
\newcommand{\dimkey}[1]{\tikz[baseline=-0.6ex]\filldraw[fill=#1, draw=#1!70!black, line width=0.4pt] (0,-0.09) rectangle (0.30,0.11);}
{\setlength{\fboxsep}{2.5pt}\setlength{\fboxrule}{0.4pt}%
\fcolorbox{black!70}{white}{\scriptsize
  \dimkey{DimWho}~Who \hspace{0.22cm}
  \dimkey{DimWhat}~What \hspace{0.22cm}
  \dimkey{DimWhen}~When \hspace{0.22cm}
  \tikz[baseline=-0.6ex]\draw[black!85, line width=0.75pt] (0,0) -- (0.35,0);~rank profile}}

\vspace{0.8mm}
\begin{tikzpicture}
\begin{axis}[
  scale only axis,
  width=0.88\columnwidth,
  height=5.4cm,
  xmin=-0.12, xmax=8.42,
  ymin=0.55, ymax=9.05,
  hide x axis,
  axis y line*=left,
  y axis line style={draw=none},
  xtick=\empty,
  ytick={\yCS,\yCJ,\yPF,\yPR,\yGC,\yUS,\yDS},
  yticklabels={{Customer\\Segments}, {Customer\\Jobs}, {Product\\Features},
               {Pain\\Relievers}, {Gain\\Creators}, {Usage\\Scenarios}, {Decision\\Stages}},
  yticklabel style={font=\scriptsize, align=right, anchor=east, inner sep=2pt,
    execute at begin node={\baselineskip=7.5pt}},
  ytick style={draw=none},
  clip=false,
]
\pgfplotsinvokeforeach{\yCS,\yCJ,\yPF,\yPR,\yGC,\yUS,\yDS}{%
  \draw[black!10, line width=0.3pt] (axis cs:-0.12,#1) -- (axis cs:8.42,#1);}

\ranklane{0}{2}{4}{6}{7}{5}{3}{1}
\ranklane{1}{1}{3}{7}{6}{5}{2}{4}
\ranklane{2}{1}{2}{6}{5}{3}{7}{4}
\ranklane{3}{4}{1}{6}{2}{5}{7}{3}

\ranklane{4.4}{1}{4}{5}{7}{6}{2}{3}
\ranklane{5.4}{2}{7}{3}{6}{4}{1}{5}
\ranklane{6.4}{2}{5}{1}{3}{4}{7}{6}
\ranklane{7.4}{3}{1}{7}{5}{2}{6}{4}

\lanelabel{0}{OpenAI\\Emb.}
\lanelabel{1}{Qwen\\Emb.}
\lanelabel{2}{GPT-5.6\\Luna}
\lanelabel{3}{DeepSeek\\4.1-flash}
\lanelabel{4.4}{OpenAI\\Emb.}
\lanelabel{5.4}{Qwen\\Emb.}
\lanelabel{6.4}{GPT-5.6\\Luna}
\lanelabel{7.4}{DeepSeek\\4.1-flash}

\draw[black!70, line width=0.45pt]
  (axis cs:0,9.05) -- (axis cs:0,9.25) -- (axis cs:3.8,9.25)
  node[pos=0.5, font=\scriptsize, text=black, fill=white, inner xsep=3pt, inner ysep=0.5pt]
    {\textbf{GPT-6-Luna}}
  -- (axis cs:3.8,9.05);
\draw[black!70, line width=0.45pt]
  (axis cs:4.4,9.05) -- (axis cs:4.4,9.25) -- (axis cs:8.2,9.25)
  node[pos=0.5, font=\scriptsize, text=black, fill=white, inner xsep=3pt, inner ysep=0.5pt]
    {\textbf{Gemini-3.1-flash-lite}}
  -- (axis cs:8.2,9.05);

\end{axis}
\end{tikzpicture}
\vspace{-4mm}
\caption{Relative rankings of seven dimension-specific strategies across eight generation--ranking configurations, based on average $\Delta$ Win gains over five datasets. Longer bars indicate better relative rankings, with rank 1 being the best. Colors distinguish What, Who, and When dimensions.}
\label{fig:rq2-config-wins}
\vspace{-2mm}
\end{figure}

\stitle{Strategy Rankings under $\Delta$ Win for RQ2}
Figure~\ref{fig:rq2-config-wins} presents the relative rankings of seven dimension-specific strategies under $\Delta$ Win, complementing the $\Delta$ Rank results in Figure~\ref{fig:rq2-config-ranks}. The top-ranked strategy is identical under both metrics in only four of the eight configurations.
Under $\Delta$ Win, Who ranks first among dimensions in three of the four LLM-based model combinations, while embedding rankers exhibit a mixture of Who and When preferences. Individual strategy rankings also vary with the generator. For example, with Qwen embeddings, switching from GPT-6 to Gemini changes the top-ranked strategy from Customer Segments to Usage Scenarios. These results show that the preferred information strategy depends on the ranking configuration and evaluation metric, complementing the configuration-specific differences reported in RQ2.

\stitle{Sensitivity to Ranking Weight $\lambda$}
We examine the sensitivity of Unified to the ranking weight $\lambda$, averaging results across five datasets and two generation models. Table~\ref{tab:lambda} shows positive average $\Delta$ Rank and $\Delta$ Win gains for both embedding rankers at all evaluated weights. Gains increase as $\lambda$ decreases, indicating that ranking improvements are sensitive to the relative weight assigned to context-hint similarity. For OpenAI embeddings, reducing $\lambda$ from 0.9 to 0.1 increases $\Delta$ Rank from 0.25 to 1.69 positions and $\Delta$ Win from 1.35 to 20.14 percentage points. Qwen embeddings show the same trend, with gains increasing from 0.28 to 2.19 positions and from 0.73 to 24.03 percentage points. We use $\lambda=0.5$ to assign equal weight to ad-card and context-hint similarities, rather than selecting the setting with the largest observed gains.

\begin{table}[!t]
\centering
\small
\vspace{-2mm}
\caption{Sensitivity of Unified to the ranking weight $\lambda$ across two embedding rankers. Bold indicates the default setting.}
\label{tab:lambda}
\vspace{-2mm}
\setlength{\tabcolsep}{5pt}
\begin{tabular}{lcccc}
\toprule
 & \multicolumn{2}{c}{OpenAI Emb.} & \multicolumn{2}{c}{Qwen Emb.} \\
\cmidrule(lr){2-3}\cmidrule(lr){4-5}
$\lambda$ & $\Delta$ Rank & $\Delta$ Win & $\Delta$ Rank & $\Delta$ Win \\
\midrule
0.1 & +1.69 & +20.14 & +2.19 & +24.03 \\
0.3 & +1.44 & +14.53 & +1.84 & +16.54 \\
\textbf{0.5} & \textbf{+1.11} & \textbf{+9.04} & \textbf{+1.40} & \textbf{+8.93} \\
0.7 & +0.71 & +4.57 & +0.87 & +3.75 \\
0.9 & +0.25 & +1.35 & +0.28 & +0.73 \\
\bottomrule
\end{tabular}
\vspace{-2mm}
\end{table}

\begin{table*}[t]
\centering
\caption{Full results of eight hint generation strategies across five datasets and eight generation--ranking configurations. No Hints reports baseline mean rank (R) and top-1 rate (W, \%), while other rows report $\Delta$ Rank (R, positions) and $\Delta$ Win (W, percentage points). Gray values indicate gains that are not statistically significant.}
\label{tab:appendix-full}
\vspace{-1.5mm}
\small
\setlength{\tabcolsep}{2.5pt}
\renewcommand{\arraystretch}{1.05}
\begin{tabular}{llrrrrrrrrrrrrrrrr}
\toprule
& & \multicolumn{8}{c}{\textbf{GPT-6-Luna}} & \multicolumn{8}{c}{\textbf{Gemini-3.1-flash-lite}} \\
\cmidrule(lr){3-10}\cmidrule(l){11-18}
& & \multicolumn{2}{c}{\makebox[0pt]{\begin{tabular}[b]{@{}c@{}}OpenAI\\Emb.\end{tabular}}} & \multicolumn{2}{c}{\makebox[0pt]{\begin{tabular}[b]{@{}c@{}}Qwen\\Emb.\end{tabular}}} & \multicolumn{2}{c}{\makebox[0pt]{\begin{tabular}[b]{@{}c@{}}GPT-5.6-\\Luna\end{tabular}}} & \multicolumn{2}{c}{\makebox[0pt]{\begin{tabular}[b]{@{}c@{}}DeepSeek-\\4.1-flash\end{tabular}}} & \multicolumn{2}{c}{\makebox[0pt]{\begin{tabular}[b]{@{}c@{}}OpenAI\\Emb.\end{tabular}}} & \multicolumn{2}{c}{\makebox[0pt]{\begin{tabular}[b]{@{}c@{}}Qwen\\Emb.\end{tabular}}} & \multicolumn{2}{c}{\makebox[0pt]{\begin{tabular}[b]{@{}c@{}}GPT-5.6-\\Luna\end{tabular}}} & \multicolumn{2}{c}{\makebox[0pt]{\begin{tabular}[b]{@{}c@{}}DeepSeek-\\4.1-flash\end{tabular}}} \\
\cmidrule(lr){3-4}\cmidrule(lr){5-6}\cmidrule(lr){7-8}\cmidrule(lr){9-10}\cmidrule(lr){11-12}\cmidrule(lr){13-14}\cmidrule(lr){15-16}\cmidrule(lr){17-18}
\textbf{Dataset} & \textbf{Strategy} & R & W & R & W & R & W & R & W & R & W & R & W & R & W & R & W \\
\midrule
\textbf{\adlib} & No Hints & 7.25 & 17.8 & 7.25 & 18.4 & 7.30 & 17.2 & 7.33 & 17.9 & 7.25 & 17.8 & 7.25 & 18.4 & 7.30 & 17.2 & 7.33 & 17.9 \\
& Customer Segments & 0.61 & 2.7 & 1.24 & 3.9 & 0.57 & 6.1 & 0.92 & 8.7 & 0.93 & 4.7 & 1.49 & 6.1 & 0.66 & 6.9 & 0.98 & 8.9 \\
& Customer Jobs & 0.57 & \textcolor{gray}{2.3} & 1.25 & 3.5 & 0.63 & 6.1 & 1.00 & 8.8 & 0.92 & 4.8 & 1.61 & 6.3 & 0.67 & 6.7 & 0.98 & 9.4 \\
& Product Features & 0.36 & \textcolor{gray}{1.6} & 1.10 & 3.0 & 0.51 & 5.2 & 0.86 & 8.4 & 0.85 & 4.8 & 1.56 & 6.7 & 0.67 & 6.4 & 0.98 & 8.9 \\
& Pain Relievers & 0.34 & \textcolor{gray}{1.5} & 1.07 & 2.7 & 0.54 & 5.5 & 0.87 & 8.5 & 0.75 & 3.6 & 1.50 & 6.1 & 0.68 & 6.9 & 0.98 & 9.1 \\
& Gain Creators & 0.39 & \textcolor{gray}{1.3} & 1.09 & 3.0 & 0.53 & 5.7 & 0.88 & 8.1 & 0.79 & 3.7 & 1.55 & 5.9 & 0.71 & 6.6 & 1.02 & 9.1 \\
& Usage Scenarios & 0.66 & \textcolor{gray}{2.2} & 1.31 & 3.4 & 0.54 & 5.0 & 0.90 & 8.1 & 0.94 & 4.7 & 1.56 & 5.7 & 0.64 & 6.2 & 0.98 & 8.3 \\
& Decision Stages & 0.40 & \textcolor{gray}{1.9} & 1.15 & 3.4 & 0.51 & 5.4 & 0.88 & 8.1 & 0.94 & 5.1 & 1.53 & 6.3 & 0.61 & 6.9 & 0.95 & 9.2 \\
& Unified & 0.64 & 2.8 & 1.23 & 3.5 & 0.55 & 5.7 & 0.92 & 8.5 & 0.94 & 4.3 & 1.65 & 6.6 & 0.72 & 7.0 & 1.02 & 9.6 \\
\midrule
\textbf{\adindia} & No Hints & 5.41 & 17.7 & 4.98 & 18.3 & 4.26 & 21.0 & 4.50 & 19.9 & 5.41 & 17.7 & 4.98 & 18.3 & 4.26 & 21.0 & 4.50 & 19.9 \\
& Customer Segments & 0.71 & \textcolor{gray}{9.1} & 0.87 & \textcolor{gray}{3.0} & \textcolor{gray}{0.41} & \textcolor{gray}{6.9} & 0.59 & \textcolor{gray}{8.3} & 1.11 & \textcolor{gray}{10.3} & 1.09 & \textcolor{gray}{5.3} & 0.65 & 7.3 & 0.99 & 9.7 \\
& Customer Jobs & \textcolor{gray}{0.46} & \textcolor{gray}{7.2} & 0.74 & \textcolor{gray}{3.6} & \textcolor{gray}{0.24} & 6.3 & 0.52 & \textcolor{gray}{8.3} & 0.81 & \textcolor{gray}{5.5} & 0.90 & \textcolor{gray}{2.1} & 0.46 & 6.3 & 0.79 & \textcolor{gray}{9.0} \\
& Product Features & 0.59 & \textcolor{gray}{7.4} & 0.74 & \textcolor{gray}{2.3} & \textcolor{gray}{0.23} & \textcolor{gray}{6.7} & 0.54 & \textcolor{gray}{8.4} & 0.74 & \textcolor{gray}{6.9} & 1.03 & \textcolor{gray}{5.0} & 0.48 & 7.6 & 0.71 & \textcolor{gray}{8.5} \\
& Pain Relievers & \textcolor{gray}{0.54} & \textcolor{gray}{5.9} & 0.75 & \textcolor{gray}{3.2} & \textcolor{gray}{0.24} & 6.3 & 0.57 & \textcolor{gray}{9.5} & 0.62 & \textcolor{gray}{6.1} & 0.89 & \textcolor{gray}{3.6} & \textcolor{gray}{0.31} & 6.2 & 0.60 & \textcolor{gray}{7.8} \\
& Gain Creators & 0.70 & \textcolor{gray}{8.0} & 0.84 & \textcolor{gray}{3.4} & \textcolor{gray}{0.20} & 6.8 & 0.55 & \textcolor{gray}{9.2} & 0.69 & \textcolor{gray}{6.3} & 0.99 & \textcolor{gray}{4.2} & \textcolor{gray}{0.39} & 7.0 & 0.63 & \textcolor{gray}{8.9} \\
& Usage Scenarios & \textcolor{gray}{0.62} & \textcolor{gray}{7.0} & 0.80 & \textcolor{gray}{2.5} & \textcolor{gray}{0.32} & 5.8 & 0.53 & \textcolor{gray}{7.5} & 0.93 & \textcolor{gray}{9.1} & 0.98 & \textcolor{gray}{5.9} & 0.52 & 6.5 & 0.74 & \textcolor{gray}{9.2} \\
& Decision Stages & 1.03 & \textcolor{gray}{10.3} & 0.95 & \textcolor{gray}{3.6} & \textcolor{gray}{0.14} & \textcolor{gray}{6.0} & 0.60 & 9.4 & 1.08 & \textcolor{gray}{8.0} & 0.99 & \textcolor{gray}{3.2} & \textcolor{gray}{0.43} & 6.5 & 0.70 & \textcolor{gray}{8.7} \\
& Unified & 0.93 & \textcolor{gray}{10.3} & 0.93 & \textcolor{gray}{3.4} & \textcolor{gray}{0.19} & \textcolor{gray}{5.6} & 0.58 & \textcolor{gray}{9.1} & 0.83 & \textcolor{gray}{9.0} & 1.04 & \textcolor{gray}{5.0} & \textcolor{gray}{0.54} & 6.8 & 0.84 & \textcolor{gray}{9.8} \\
\midrule
\textbf{\lmmarket} & No Hints & 5.21 & 12.9 & 5.42 & 10.9 & 5.26 & 11.4 & 5.22 & 11.5 & 5.21 & 12.9 & 5.42 & 10.9 & 5.26 & 11.4 & 5.22 & 11.5 \\
& Customer Segments & 0.97 & 8.2 & 1.73 & 13.6 & 0.27 & 3.5 & 0.38 & 6.0 & 1.51 & 14.8 & 1.92 & 15.9 & 0.30 & 5.3 & 0.44 & 6.5 \\
& Customer Jobs & 0.96 & 9.4 & 1.73 & 13.6 & 0.31 & 4.3 & 0.38 & 5.9 & 1.48 & 13.8 & 2.02 & 17.3 & 0.28 & 4.6 & 0.48 & 6.5 \\
& Product Features & 0.85 & 7.2 & 1.59 & 11.0 & 0.27 & 3.9 & 0.37 & 6.3 & 1.44 & 13.4 & 1.96 & 16.7 & 0.32 & 5.2 & 0.44 & 6.4 \\
& Pain Relievers & 0.85 & 7.8 & 1.58 & 12.1 & 0.27 & 4.7 & 0.42 & 6.1 & 1.32 & 11.8 & 1.95 & 17.0 & 0.33 & 5.2 & 0.45 & 6.7 \\
& Gain Creators & 0.91 & 8.2 & 1.62 & 11.8 & 0.32 & 4.8 & 0.42 & 6.4 & 1.34 & 11.9 & 1.99 & 16.0 & 0.29 & 4.3 & 0.48 & 6.5 \\
& Usage Scenarios & 1.05 & 9.5 & 1.76 & 13.9 & 0.31 & 4.2 & 0.39 & 5.9 & 1.55 & 14.3 & 2.01 & 16.5 & 0.28 & 5.0 & 0.44 & 6.3 \\
& Decision Stages & 1.03 & 9.7 & 1.71 & 12.9 & 0.27 & 4.5 & 0.38 & 6.3 & 1.60 & 14.6 & 1.96 & 16.4 & 0.31 & 4.4 & 0.47 & 6.4 \\
& Unified & 1.09 & 9.1 & 1.71 & 12.8 & 0.28 & 4.5 & 0.37 & 6.6 & 1.41 & 14.1 & 2.04 & 17.8 & 0.36 & 4.9 & 0.52 & 7.0 \\
\midrule
\textbf{\naiad} & No Hints & 5.42 & 9.1 & 5.44 & 7.4 & 5.38 & 10.3 & 5.47 & 10.0 & 5.42 & 9.1 & 5.44 & 7.4 & 5.38 & 10.3 & 5.47 & 10.0 \\
& Customer Segments & 1.41 & 12.3 & 1.98 & 16.0 & \textcolor{gray}{0.10} & 4.7 & 0.44 & 7.1 & 1.72 & 14.8 & 2.12 & 18.4 & 0.20 & 4.6 & 0.52 & 7.6 \\
& Customer Jobs & 1.43 & 12.5 & 1.96 & 15.2 & \textcolor{gray}{0.13} & 4.8 & 0.45 & 7.7 & 1.71 & 15.6 & 2.16 & 18.0 & 0.20 & 5.3 & 0.53 & 8.7 \\
& Product Features & 1.24 & 10.2 & 1.91 & 15.4 & \textcolor{gray}{0.07} & 3.7 & 0.40 & 6.7 & 1.63 & 13.7 & 2.12 & 17.3 & \textcolor{gray}{0.17} & 5.1 & 0.50 & 7.1 \\
& Pain Relievers & 1.17 & 8.6 & 1.86 & 13.7 & \textcolor{gray}{0.06} & 3.4 & 0.41 & 6.9 & 1.50 & 12.5 & 2.10 & 16.7 & \textcolor{gray}{0.14} & 5.0 & 0.49 & 7.2 \\
& Gain Creators & 1.33 & 10.7 & 1.88 & 14.3 & \textcolor{gray}{0.13} & 4.8 & 0.41 & 6.2 & 1.62 & 13.3 & 2.12 & 17.8 & \textcolor{gray}{0.17} & 4.7 & 0.47 & 8.2 \\
& Usage Scenarios & 1.45 & 13.0 & 2.01 & 16.0 & \textcolor{gray}{0.09} & 4.7 & 0.40 & 7.4 & 1.84 & 15.8 & 2.24 & 19.5 & 0.21 & 4.6 & 0.50 & 8.1 \\
& Decision Stages & 1.37 & 9.9 & 1.97 & 14.7 & \textcolor{gray}{0.01} & 3.6 & 0.31 & 6.3 & 1.70 & 14.8 & 2.10 & 17.0 & \textcolor{gray}{0.13} & 4.3 & 0.50 & 7.4 \\
& Unified & 1.51 & 11.2 & 2.02 & 14.7 & \textcolor{gray}{0.09} & 4.4 & 0.41 & 7.8 & 1.66 & 14.4 & 2.20 & 19.8 & 0.29 & 5.3 & 0.54 & 8.8 \\
\midrule
\textbf{\egeo} & No Hints & 5.16 & 12.6 & 5.24 & 10.8 & 5.12 & 13.2 & 5.13 & 13.2 & 5.16 & 12.5 & 5.24 & 10.8 & 5.12 & 13.2 & 5.13 & 13.2 \\
& Customer Segments & 0.79 & 5.2 & 0.27 & \textcolor{gray}{0.7} & 1.35 & 20.7 & 1.41 & 23.7 & 1.34 & 10.4 & 0.87 & 5.0 & 1.44 & 22.2 & 1.42 & 23.4 \\
& Customer Jobs & 0.64 & 4.5 & 0.18 & \textcolor{gray}{0.3} & 1.31 & 20.2 & 1.42 & 23.8 & 1.25 & 9.8 & 0.83 & 4.0 & 1.43 & 22.0 & 1.41 & 23.0 \\
& Product Features & 0.66 & 3.9 & \textcolor{gray}{0.14} & \textcolor{gray}{0.0} & 1.29 & 19.7 & 1.35 & 22.8 & 1.25 & 9.6 & 0.88 & 4.4 & 1.47 & 22.1 & 1.42 & 23.8 \\
& Pain Relievers & 0.67 & 5.0 & 0.21 & \textcolor{gray}{0.6} & 1.31 & 20.0 & 1.40 & 23.7 & 1.25 & 9.2 & 0.89 & 4.9 & 1.48 & 22.9 & 1.45 & 24.1 \\
& Gain Creators & 0.63 & 4.7 & 0.19 & \textcolor{gray}{0.0} & 1.30 & 19.6 & 1.41 & 23.4 & 1.22 & 8.9 & 0.86 & 4.8 & 1.47 & 22.6 & 1.44 & 23.8 \\
& Usage Scenarios & 0.73 & 5.1 & 0.25 & \textcolor{gray}{0.8} & 1.28 & 19.4 & 1.37 & 22.9 & 1.22 & 9.7 & 0.77 & 3.8 & 1.36 & 20.8 & 1.37 & 22.8 \\
& Decision Stages & 0.83 & 6.0 & 0.40 & \textcolor{gray}{1.4} & 1.34 & 20.7 & 1.43 & 24.1 & 1.36 & 10.8 & 0.93 & 5.4 & 1.48 & 22.8 & 1.47 & 24.0 \\
& Unified & 0.85 & 5.7 & 0.35 & \textcolor{gray}{1.1} & 1.32 & 20.6 & 1.43 & 24.4 & 1.22 & 9.4 & 0.89 & 4.6 & 1.53 & 23.6 & 1.52 & 25.2 \\
\bottomrule
\end{tabular}
\vspace{-1mm}
\end{table*}

\end{document}